\documentclass[prx,superscriptaddress,twocolumn,twoside,floatfix,pra,a4paper,nofootinbib]{revtex4-2} 
\usepackage{times}
\usepackage{epsfig}
\usepackage{amsfonts}
\usepackage{amsmath}
\usepackage{array}
\usepackage{amssymb,amsthm}
\usepackage{color}
\usepackage{multirow}
\usepackage{braket}
\usepackage{bbm}
\usepackage{latexsym}
\usepackage{amsfonts}
\usepackage{mathrsfs}
\usepackage{natbib}
\usepackage{verbatim}
\usepackage{gensymb}
\usepackage{caption}
\usepackage{subcaption}
\usepackage{subcaption}
\usepackage{graphicx}
\usepackage{ dsfont }
\usepackage[colorlinks=true,linkcolor=blue,citecolor=magenta,urlcolor=blue]{hyperref}
\usepackage[qm]{qcircuit}
\allowdisplaybreaks

\usepackage{multirow}

\begin{document}

\newcommand{\qc}[1]{Circuit~(\hyperref[qc:#1]{\ref*{qc:#1}})}
\newcounter{qcnum}
\newcommand{\qcref}[1]{\refstepcounter{qcnum}\tag{\theqcnum}\label{qc:#1}}


\title{Nonclassicality of a Macroscopic Qubit-Ensemble via Parity Measurement Induced Disturbance}

\author{Lorenzo Braccini}
\email{lorenzo.braccini.18@ucl.ac.uk}
\affiliation{Department of Physics and Astronomy, University College London, Gower Street, London WC1E 6BT, England, United Kingdom}

\author{Debarshi Das}
\email{dasdebarshi90@gmail.com}
\affiliation{Department of Physics, Shiv Nadar Institution of Eminence, Gautam Buddha Nagar, Uttar Pradesh 201314, India}

\author{Ben Zindorf}
\affiliation{Department of Physics and Astronomy, University College London, Gower Street, London WC1E 6BT, England, United Kingdom}


\author{Stephen D. Hogan}
\affiliation{Department of Physics and Astronomy, University College London, Gower Street, London WC1E 6BT, England, United Kingdom}

\author{John J. L. Morton}
\affiliation{London Centre for Nanotechnology, University College London, 17-19 Gordon Street, London WCH1 0AH, England, United Kingdom}
\affiliation{Department of Electronic \& Electrical Engineering, University College London, London
WC1E 7JE, England, United Kingdom}

\author{Sougato Bose}
\affiliation{Department of Physics and Astronomy, University College London, Gower Street, London WC1E 6BT, England, United Kingdom}
	
\begin{abstract}

We propose an experimental scheme to test the nonclassicality of a macroscopic ensemble of qubits, through the violation of the classical notion of macrorealism (MR) via the fundamental measurement-induced disturbance of quantum systems. An electromagnetic resonator is used to probe the parity of the qubit-ensemble. The action of sequential measurements allows the nonclassicality of whole ensemble to manifest itself, in the ideal case, irrespective of its size. This enables to probe the macroscopic limits of quantum mechanics as the qubit-ensemble is, effectively, a single large spin of many $\hbar$ units.  Even as $\hbar \rightarrow 0$ in comparison to the total angular momentum of the ensemble, a constant amount of violation of MR is found in the noiseless case. However, environmental decoherence and inhomogeneity of qubit-electromagnetic field couplings precipitate the quantum-to-classical transition. This illustrates that Bohr's correspondence principle is not fundamental, but a consequence of practical limitations. We outline an implementation with a variety of qubits (superconducting qubits, spins in semiconductors, and Rydberg atoms) coupled to a coplanar waveguide resonator, and -- via the corresponding noise analysis -- find that violation of MR is detectable up to $100$ qubits via current technology.

\end{abstract}
\maketitle

\section{Introduction}

Probing the macroscopic limits of quantum mechanics, and thereof the boundary between quantum and classical physics, is one of the biggest open quests in modern science. Only persistent experimental efforts to observe genuine quantum mechanical phenomena in more and more macroscopic systems can address this issue. While spin is an important parameter to characterize the `macroscopicity' of a system (i.e., an object with large spin can be considered as macroscopic), experiments to test the quantum nature of a large spin are still lacking. As opposed to other parameters, such as mass or spatial spread, spin is the only quantity directly comparable to the action in terms of $\hbar$, which has been historically used for explaining quantum-to-classical transition: $\hbar \rightarrow 0$ is traditionally believed to imply the classical limit of quantum mechanics. Although it is now well understood that the emergence of classicality generally results from ubiquitous interactions with the environment, and maintaining sufficient isolation from the environment becomes increasingly challenging in the $\hbar \rightarrow 0$ limit. For example, a macroscopic magnet, which can be approximated as a collection of qubit spins, behaves, in general, as a large classical vector. Against this backdrop, we propose a realizable experiment to test the quantumness of a macroscopic ensemble of qubits, simulating a large spin. Specifically, our aim is to address the following: \textit{When does a large ensemble of qubits, with a total spin of many units of $\hbar$, stop behaving as a nonclassical entity?} 


According to our protocol, in the ideal case, there is no such fundamental limit. However, experimental imperfections inevitably impose practical constraints on the ensemble size. Despite this, our proposed scheme can be implemented using different types of physical qubit-ensembles, and with current technology, it is capable of demonstrating nonclassicality with up to approximately $50-100$ qubits -- equivalent to a spin of $\sim 50 \hbar - 25 \hbar$. This opens a new avenue for testing genuine nonclassicality in large spin systems, extending into a macroscopic regime hitherto unexplored. Notably, this approach enables a simpler test of nonclassicality with spins of up to an order of magnitude greater than those of previous experiments, which had been limited to values below 10 \cite{Athalye_2011,knee2012violation,newton}.

As a tool for testing quantumness, we use the violation of the classical notion of macrorealism (MR)~\cite{lgi1,leggett02,lgi2}. MR is the conjunction of the two assumptions: (1) \textit{Realism per se:} At any instant, even if unobserved, a system is definitely in one of its possible states with all its observable properties having definite values. (2) \textit{Noninvasive measurability:} It is possible to determine which of the states the system is in by ensuring the measurement induced disturbance is arbitrarily small, thus not affecting the state or the subsequent time evolution of the system~\cite{lgi1}. Quantum systems violate these two assumptions due to the superposition principle and the measurement induced wavefunction collapse, respectively. From the above two assumptions, various necessary testable conditions of MR can be derived, for instance, the Leggett-Garg inequalities~\cite{lgi1,qlgi1} and the No Disturbance Condition (NDC)\footnote{also known as the no signalling in time condition \cite{nsit1}.}~\cite{nsit1,nsit2,nsit3,largemass,gravity_measurement}, with the latter being used in this work. Experimental refutation of any of these conditions in a loophole-free way implies the inherent nonclassicality or quantumness of a measured system. 

Large qubit-ensembles have been established as reliable quantum resources for computing~\cite{computing1,memory1,Chen2018,computing2,qubit_ensemble_computing_1}, sensing~\cite{sensing1,sensing2,metro}, memory~\cite{morton_memory_1,memory2,mem,mem1,memory3,mem4,mem2,memory4,memory5,mem3,mmry,memory6,memory7,memory8,memory9,memory10, memory11}, and communication~\cite{comm,comm1,comm2}. Despite all these advances, its quantum property as a large collective spin remains unexplored, since usually these studies are limited to the low-excitation sector, where the system can be {\em approximated} to a quantum harmonic oscillator (through the Holstein–Primakoff transformation~\cite{HP_transformation}).
On the other hand, experimental violation of MR has been reported earlier only for single qubit or \textit{microscopic} spin systems~\cite{Athalye_2011,knee2012violation,sadana2022testing,joarder2022loophole,sadana2023noise,nath2024single}. Although large spins have been considered in literature for probing quantumness~\cite{lsnew1,lsnew2,lsnew3,lsnew4,large_spin1,large_spin2,lsnew5,large_spin3,large_spin4,large_spin5,large_spin6,large_spin7}, all these analyses are purely theoretical with no scope for near-term experimental realization. While quantum effects such as tunneling, squeezing, resonant transitions, coherence and precession have been experimentally observed in large-spin systems~\cite{newton,s9,s10-1,s10-2,s150-1,s150-2,s150-3,BEC,Nucleus,qi2}, these demonstrations neither test the quantumness via violation of MR, nor can easily scale to macroscopic levels. 

Here, we close the gap between the current experiments involving qubit-ensembles -- which have had mainly practical applications 
-- and the above foundational questions of quantum mechanics of large spins -- which, until now, have been lacking any realisable experimental schemes. This is achieved by exploiting the dispersive interaction between an ensemble of qubits and a common resonator, alongside homodyne measurements. This interaction has been central for entanglement gates between qubits in numerous architectures, known as strong coupling~\cite{Xiang2013,PhysSuper1,PhysSuper2,PhysSuper3,Phys2Super1,PhysSolid1,PhysSolid2,PhysSolid3,Pritchard2014,PhysRygberHogan,rydberg_array}. Moreover, measurement-based entanglement of two qubits, via this interaction with inclusion of homodyne measurement, has been demonstrated in superconducting and solid state experiments~\cite{Ruskov2003,Lalumiere2010,Roch2014,PhysSolidMeasureBased,delva2024}. Under specific parameter choices, we show that the generalisation of these protocols to $N$ qubits can be used to perform parity measurements of the ensemble. Thus, an ensemble of $N$ qubits can violate MR through NDC by performing two such consecutive parity measurements.

We also note recent proposals for macroscopic MR tests based on quantum nondemolition measurements, which carefully distinguish unavoidable quantum measurement backaction from incidental classical disturbances (see details in Sec. \ref{sec:classical_disturbance}) \cite{loop2,loop4,loop5}. The present work complements these approaches by developing a collective parity measurement-based protocol, analysing its experimental implementation and robustness against realistic imperfections.

In Sec.~\ref{sec:ideal_case}, the ideal protocol is proposed, for which a constant quantum violation of MR is detectable with arbitrary number of qubits. In Sec.~\ref{sec:quantum_classical}, we find that a quantum-to-classical transition for large spins is induced by practical constraints, namely, operational imperfection, decoherence, and inhomogeneities among qubits. In Sec.~\ref{sec:classical_disturbance}, the experimental challenge of avoiding classical disturbances in NDC detection is discussed. In Sec.~\ref{sec:physical_system}, the experimental requirements for testing MR with different qubit-ensembles -- Rydberg atoms, quantum dot spins, and superconducting qubits -- are presented, where bounds on the maximum number of qubits for which the violation of MR is detectable are derived given the current state-of-the-art in \textit{noisy} technologies, concluding that $N_{\text{NDC}} \sim 50 -100$ qubits. In our parallel work~\cite{upcoming}, the proposed protocol is mapped to gates and used on IBM Quantum Computers (QCs) to detect non-classicality up to $38$ qubits\footnote{increasing by one order of magnitude the best known results of detection of violation of MR on QCs~\cite{LGIQC1,LGIQC2,LGIQC3,LGIQC4,LGIQC5}.}.


\section{Ideal Protocol \label{sec:ideal_case}}

In this Section, we begin by introducing the form of the two-time NDC and collective qubits-light interactions, used in this work. Then, the proposed protocol is presented for the ideal case, which is with infinite operational precision and in the absence of noise and decoherence. The analysis is first kept general for an ensemble of $N$ qubits that interacts with a common resonator or cavity, as such system can be physically implemented in numerous implementations (see Sec. \ref{sec:physical_system}). 


\subsection{The No Disturbance Condition (NDC) \label{sec:no_siturbance}}

We begin by introducing the form of the two-time NDC that involves sequential measurements of a dichotomic observable (with $\pm$ outcomes) at two different instants $t_1$ and $t_2$, where $t_1 < t_2$. The NDC implies that the probability of
obtaining a particular outcome for the measurement at $t_2$ should be independent of whether a previous measurement has been performed. Mathematically, the NDC can be expressed as
\begin{align}
	&V_{\pm}  =	P_2(\pm) - \left[  P_{1,2}(+,  \pm) +  P_{1,2}(-,\pm) \right] =0.
	\label{eq:ndc}
\end{align}
where, for instance, $P_{1,2}(-, +)$  is the joint probability of getting the outcomes $-$ at instant $t_1$ and $+$ at instant $t_2$, and $P_{2}(+)$ is the probability of outcome $+$ at instant $t_2$ without any measurement at $t_1$. If classical disturbance is not present (see Sec.~\ref{sec:classical_disturbance}), any non-zero value of $V_{\pm}$ denotes quantum violation of the NDC, as the system wavefunction undergoes the measurement induced non-unitary collapse. Here we note that from normalization of probabilities, it follows that $V_+ = - V_-$, from which we will be interested in the magnitude of such quantity $|V_{\pm}|$.

\subsection{Collective Qubit-Light Interactions}

Let us consider $N$ qubits, with Pauli operators $\sigma_{\alpha}^{(i)}$ (where $i \in [1, N]$ labels the qubit and $\alpha = x,y,z$, from which $\sigma_\pm = \sigma_x \pm i \sigma_y$), and a resonator or cavity of frequency $\omega$ with annihilation and creation operators being $a$ and $a^\dag$. All the Hamiltonians are given in units of $\hbar$, i.e., the coupling is in the angular frequency units. In the interaction picture, the rotating-wave approximation is considered, and the qubit-cavity interaction is given by the Jaynes-Cummings interaction Hamiltonian for $N$-qubits
\begin{equation}
\label{eq:hamiltonian_reson}
    H_{\text{JC}} =   \sum_{i=1}^N  g_i (a \sigma_+^{(i)} + a^\dag \sigma_-^{(i)}) \;,
\end{equation}
where $g_i$ is the coupling between the $i$-th qubit and the cavity~\cite{JC_physics}. 

As first analysis, the qubits are assumed to be identical and uniformly interacting with the cavity: the former assumption implies that all the qubits have the same natural frequency $\omega_0$ and the latter ensures that the qubit-cavity coupling constant is the same for all qubits, i.e. $\{g_i = g \; \forall i \in [0,N] \}$. We will further assume that there is no qubit-qubit interaction (which physically can be ensured by spacing the qubits at large enough distances, as these interactions usually decay rapidly with the distance, for instance, the dipole-dipole magnetic interaction, see Sec.~\ref{sec:physical_system}). Under the above assumption, the qubit-ensemble can be treated as a single large spin with angular momentum $j = N/2$, Hilbert space $\mathcal{H} \simeq \mathds{C}^{2j+1}$, and spin operators $J_{\alpha} := (1/2)\sum_{i=0}^{N} \sigma_{\alpha}^{(i)}$,  with $\alpha = x,y,z,+,-$~\cite{dicke_coherence_1954}. Note that this approximation is routinely done theoretically and can be satisfied in experimental scenarios ~\cite{schliemann_coherent_2015, wang_giant_2022, ma_quantum_2011}. Under these assumption, the Hamiltonian of Eq.~(\ref{eq:hamiltonian_reson}) reduces to $\approx 2  g (a J_+ + a^\dag J_-)$, which is known as the Tavis-Cummings model~\cite{travis}. The detuning frequency between the cavity and the qubits is defined as $\Delta := \omega - \omega_0$. In the resonance regime ($\Delta \approx 0$), the cavity can be used to rotate the spin and for state preparation. With a cavity field in a Coherent State (CS)\footnote{A CS is defined as the quantum state describing the displacement vacuum $\ket{\alpha} = e^{\alpha a^\dag - \alpha^* a} \ket{0} $ with $\alpha = \in \mathbb{C}$. A CS is the eigenstate of the annihilation operator, $a \ket{\alpha} = \alpha \ket{\alpha}$, and, in the $n$-basis representation, takes the form
$\Ket{\alpha} = e^{- \frac{1}{2} |\alpha|^2} \sum_n (\alpha)^n/\sqrt{n!}  \Ket{n}$. They form an overcomplete basis.} $\Ket{\alpha_r}_c$ with $\alpha_r \in \mathbb{C}$ and $|\alpha_r| \gg 0$, the cavity is a classical pulse, which can absorbed (and emitted) by the qubit. In fact, it is possible to approximate $a^\dag \ket{\alpha_r}_c \approx \alpha^*_r \ket{\alpha_r}_c$ (and $a \ket{\alpha_r}_c \approx \alpha_r \ket{\alpha_r}_c$), from definition, such that the Hamiltonian effectively acts on the spin as a rotation, i.e. $ \braket{ \alpha_r | H_{\text{JC}} | \alpha_r } \approx - 4 |\alpha_r| g J_y$, where the cavity was traced out. In the off-resonance regime ($\omega \gg \Delta \gg g$), the interaction Hamiltonian can be used to entangle the qubit-ensemble with the cavity field in order to perform measurement on the spin. The effective Hamiltonian (from the second order term in the Dyson series, see Appendix~\ref{app:effective}) is 
\begin{align}
    H_{\text{E}} &=  \sum_{i=1}^N \frac{2 g_i^2 }{ \Delta } a^{\dag} a \sigma_z^{(i)} + \sum_{i\neq j}^N \frac{g_i g_j}{ \Delta }  \left( \sigma^{(i)}_+  \sigma^{(j)}_- +  \sigma^{(i)}_-  \sigma^{(j)}_+ \right) \nonumber \\ 
    &\approx \chi a^{\dag} a  J_z + 2 \chi \mathbf{J}^2 - 2 \chi J_z^2  \;,
    \label{eq:hamiltonian_off}
\end{align}
where $\chi = \frac{4 g^2 }{ \Delta } $ is the dispersive coupling for the total spin.

\subsection{Proposed Protocol}
\label{subsec:ideal}

The proposed protocol to measure the NDC using collective qubits-light interactions achieved via five stages: (i) state initiation, (ii) first rotation, (iii) first measurement, (iv) second rotation, and (v) final measurement.
This protocol is the same for integer and half-integer spin-$j$, i.e. even and odd $N$ qubits. Where else not stated, $j$ is considered to be an integer, and the half-integer case is covered in the Appendix~\ref{app:analytical_section}, where the mathematical details and proofs are also presented. In the following, quantum states with the subscript $c$ belongs to the cavity Hilbert space, while $s$ to the spin Hilbert space. When both subscripts appear, they represent a joint state in the spin-cavity Hilbert space.

(i) \textit{System Initiation:} The spin-$j$ system describing the qubit-ensemble is initially prepared in the ground state $\ket{j,m=-j}_s = \ket{1/2, -1/2}_s^{\otimes N}$, where $\ket{j,m}_s$ is the eigenstate of $J_z$ with eigenvalue $m$ associated with a spin-$j$ system.

(ii) \textit{First Rotation:} The spin interacts for a time $t_{\text{R}}$ with the cavity in resonance. The cavity field at this step is in a CS $\Ket{\alpha_r}_c$ with $\alpha_r \in \mathbb{R}$ and $\alpha_r \gg 0$. The resulting unitary evolution is a rotation $\mathcal{R}(\phi) = e^{i \phi J_y}$, where $\phi = 2 \alpha_r g t_{\text{R}} $. After the interaction the spin state is a CS defined as
\begin{align}
\label{eq:spin_cs}
  \Ket{\Psi (t_R)}_{s} &=  \Ket{\phi}_s \nonumber := \mathcal{R}(\phi) \ket{j,-j}_s \nonumber \\
  &= \sum_{m= -j}^{j} d^{(j)}_{m,  -j} \left(\phi \right) \ket{j,m}_s,
\end{align}
where $d^{(j)}_{m',  m} \left(\phi \right)$ are the Wigner d-matrix elements. 

\begin{figure}[t!]
    \centering
 \includegraphics[width=0.49\textwidth]{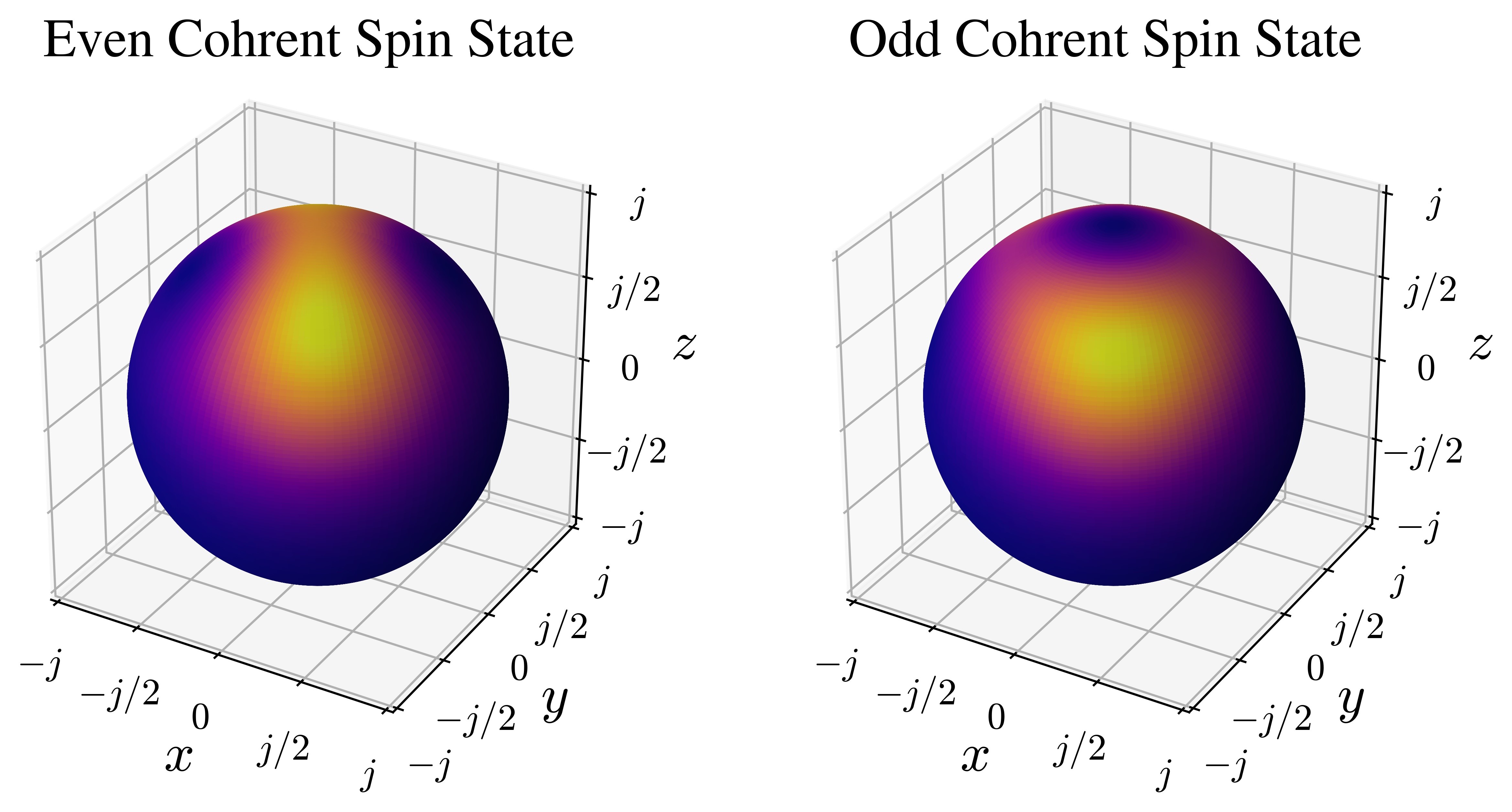}
    \caption{Husimi representations of even and odd CS of a large spin ($j=10$).}
    \label{fig:quasi_probability}
\end{figure}

(iii) \textit{First Measurement:} The measurement procedure consists of firstly entangling the system (large spin) with the probe (cavity field), followed by a homodyne detection of the cavity. The cavity in the off-resonance regime is prepared in a CS $\Ket{\alpha_0}_c$ with $\alpha_0 = \frac{(1+i)}{\sqrt{2}} |\alpha_0|$ and $|\alpha_0| \gg 1$. The choice of complex parameter $\alpha_0 $ ensures violation of NDC with integer as well as half-integer spins (see Appendix~\ref{app:analytical_section}). The interaction time $t_I$ is chosen such that $\chi t_I= \pi$. The spin-cavity entangled state is 
\begin{equation}
\label{eq:cavity_spin_entnagled}
    \Ket{\Psi(t_R + t_I)}_{sc} = \ket{\phi_e}_{s} \Ket{ \alpha_0}_c + \ket{\phi_o}_{s} \Ket{- \alpha_0}_c,
\end{equation}
where the unnormalized even and odd spin CS are defined as
\begin{align}
\label{first_interaction_state}
    & \ket{\phi_e}_{s}   :=  \sum_{m \in \text{even}}d^{(j)}_{m, -j} \left( \phi \right) \ket{j, m}_s \;, \\  & \ket{\phi_o}_{s} :=  \sum_{m \in \text{odd}}d^{(j)}_{m, -j} \left( \phi \right)  \ket{j, m}_s \;,
\end{align}
respectively. Even and odd CSs are orthogonal and posses interesting non-Gaussianity -- hence, nonclassicality -- in the Husimi-Q quasiprobability distribution (shown in Fig.~\ref{fig:quasi_probability}). This quasiprobability distribution is a ``phase-space'' representation for large spins and is defined for the spin state $\ket{\psi}_s$  according to $Q(\phi, \theta) := \left| \braket{\phi,\theta| \psi}_s \right|^2 $, where $\ket{\theta,\phi}_S$ is a spin CS with additional phase $\theta$.

Subsequently, a measurement with outcomes $+$ and $-$ defined by the following projectors is performed on the cavity field:
\begin{equation}
\label{projectionsmain}
    \Pi_+ = \int_0^{\infty} dx \Ket{x}\bra{x}_c, \;\;\;\;\;\;\;  
    \Pi_- = \int_{-\infty}^0 dx \Ket{x}\bra{x}_c.
\end{equation} 
This measurement can be approximately realized by a homodyne measurement in the limit of high intensity \textit{local oscillator}, which is high intensity coherent state of the reference signal~\cite{ferraro_gaussian_2005}.

Assuming $|\alpha_0| \gg 1$ and tracing out the cavity degrees of freedom, the unnormalized (labeled by the superscript U) post-measurement states of the spin, when the outcomes $+$ and $-$ are detected, are given by $\ket{\Psi^+}^{\text{U}}_{s} = \ket{\phi_e}_{s}$ and $\ket{\Psi^-}^{\text{U}}_{\text{s}} = \ket{\phi_o}_{s}$, respectively. Thus, the above entangling interaction between the spin (target system) and the cavity field (probe) along with the projective measurement on the cavity is effectively a dichotomic parity measurement on the spin system, described by the two projectors:
\begin{equation}
    \mathcal{S}_e = \sum_{m \in \text{even}} \ket{j,m} \bra{j,m}_s \;, \hspace{0.3cm}  
    \mathcal{S}_o = \sum_{m \in \text{odd}} \ket{j,m} \bra{j,m}_s   \;.
    \label{projector_first_measurment}
\end{equation}

By exploiting the algebraic structure of even and odd spin CS (see Sec.~\ref{app:analytical_section}), the outcome probabilities of this first measurement are given by the norms of the even and odd CS, which are analytically computed to be
\begin{equation}
\label{eq:prob_1_ideal}
 P_1(\pm) \approx  \frac{1}{2} \pm \frac{1}{2}  \left( \cos \phi \right)^{2j}.
\end{equation}

(iv) \textit{Second Rotation}:  The unnormalized states after the first measurement can be expressed as 
\begin{equation}
\label{eq:UPM_first_ideal}
\Ket{\Psi^{\pm}}^{\text{U}}_{s} = \nonumber   \frac{1}{2} \left( \Ket{\phi}_s \pm \Ket{-\phi}_s  \right),
\end{equation}
if $j$ is even integer (the other cases are presented in Appendix~\ref{app:analytical_section}). Subsequently, the system is again subjected to a second rotation by the same angle $\phi$. The spin state after the second rotation is
\begin{equation}
\label{eq:UPM_first_ideal}
\Ket{\Psi^{\pm} (t_R)}^{\text{U}}_{s} = \frac{1}{2} \left( \Ket{2 \phi}_s \pm \Ket{j,-j}_s  \right),
\end{equation}

(v) \textit{Final measurement}: Following, a measurement as described in step (iii) is performed again (by introducing a cavity field in the same state $|\alpha_0\rangle_c$, entangling it with the spin and performing the aforementioned projective measurement on the cavity). Following a similar calculation, the outcome statistics can be obtained, i.e. $P_{1,2}(+,+)$, $P_{1,2}(+,-)$, $P_{1,2}(-,+)$, $P_{1,2}(-,-)$ can be analytically derived. 

In order to determine $P_{2}(\pm)$, the first measurement is not performed, i.e.,  the step (iii) described above is not carried out. Effectively, the spin system is rotated by an angle $2 \phi$. Following, the off-resonance interaction and the aforementioned projective measurement on the cavity leads to $P_{2}(\pm)$, given by Eq.~(\ref{eq:prob_1_ideal}) with $\phi \to 2 \phi$.

In the above scenario, the violation of the NDC is computed analytically. For (even and odd) integers $j$, the violations are found to be
\begin{equation}
\label{eq:violation_ideal}
    V_{\pm}^{j = \text{even}} =\mp \frac{1}{4} \left[1- (\cos  2 \phi )^{2j} \right] = V_{\mp}^{j = \text{odd}}.
\end{equation}
The case of half-integer spin -- that is, when the ensemble is formed by an odd number $N$ of qubits -- is solved in Appendix~\ref{app:analytical_section} by rewriting $j = n + \frac{1}{2}$ (where $n$ is integer and $N = 2 n + 1$). For $n$ even (odd), the violation recovers the same expression of odd (even) integer $j$.  

\begin{figure}[t!]
    \centering
    \includegraphics[width=0.45\textwidth]{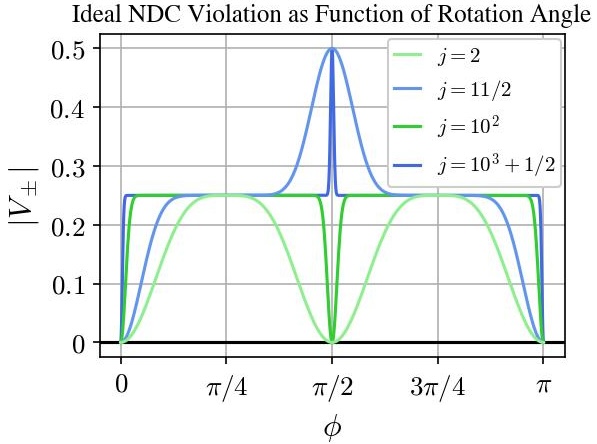}
    \caption{Variation of the magnitude of violation of the NDC, denoted by $|V_{\pm}|$, versus $\phi$ for different $j$ in the ideal case.}
    \label{plot_ideal_1}
\end{figure}

Fig.~\ref{plot_ideal_1} shows the violations as a function of angle $\phi$ for different spin values $j$. The maximum violation for integer $j$ is $|V_{\pm}| = 1/4$, whereas that for half-integer $j$ is $|V_{\pm}| = 1/2$. Interestingly, the range of $\phi$ for which $|V_{\pm}| = 1/4$ becomes broader, as $j$ increases. However, for half-integer spin, the range of $\phi$ for which one can get violations $|V_{\pm}| > 1/4$ becomes narrower with an increase in~$j$. In an experimental setting, the value of $j$  -- i.e., the number of qubits in the ensemble -- may be unknown. Since $|V_{\pm}| = 1/4$ for $\phi = \pi/4$ with any spin value, this angle is the preferred choice.

\section{The Quantum-Classical Transition \\ \& Noise Analysis \label{sec:quantum_classical}}

In the ideal case, a constant amount of violation of MR has been analytically found for any ensemble size with $\phi=\pi/4$.  Hence, a macroscopic qubit-ensemble, independent of its size, would show quantum behaviour under the presented protocol. In this section, we investigate the emergence of the classicality from quantum mechanical description~\cite{large_spin1,large_spin3,large_spin5,schlosshauer_quantum_2007,classicality_emergence}, in the context of the system studied. As we shall show, (a) experimental imperfections, (b) decoherence, and (c) inhomogeneity in the qubit-cavity couplings cause the quantum-to-classical transition of the system under consideration, with NDC tending to zero as the noise increases. 

\subsection{Rotation Errors}

Let us first consider imperfections in the rotation angles. Such error can arise due to (i) a systematic error in the coupling $g$, or (ii) error in the rotation time $t_r$, or (iii) error in the cavity intensity $\alpha_r$. Regarding the first, the precise value of the mean coupling $g$ may be unknown in an experimental setting, resulting in two consecutive equal rotations, with $\phi \neq \pi/4$. As shown in Fig.~\ref{plot_ideal_1} and Eq.~(\ref{eq:violation_ideal}), the violation of NDC is approximately constant near $\phi = \pi/4$ (and the range of $\phi$ near $\phi = \pi/4$, for which $|V_{\pm}| \approx 1/4$,  \textit{increases} as $j$ increases). Hence, by choosing $\phi = \pi/4$, the protocol is robust under this source of error. As shown in Fig~\ref{plot_ideal_1}, by measuring at $\phi = \pi/2$, one may infer the parity of number of qubits as the violation reaches a minimum (maximum) for even (odd) $N$. However, this first type of uncertainty ensures that it is not possible to distinguish between an even and odd $N$ of an arbitrarily large macroscopic ensemble in realistic experimental scenarios, as the region of $\phi$ where the violations between even/odd $N$ are distinguishable, decreases with increasing $j$. 

For the latter two errors, if the input state of the cavity field $|\alpha_r \rangle$ and/or the rotation time differ between the two rotations, the resulting rotations would have two distinct angles $\phi_1, \; \phi_2$. The violation of NDC for two arbitrary rotation angles is derived in Appendix~\ref{app2} and plotted in Fig.~\ref{fig:phi121}. The ideal case of $ \phi_1 = \phi_2$ results in the maximum violation of NDC. 
\begin{figure}[t!]
      \centering
      \subfloat{\includegraphics[width=0.40\textwidth]{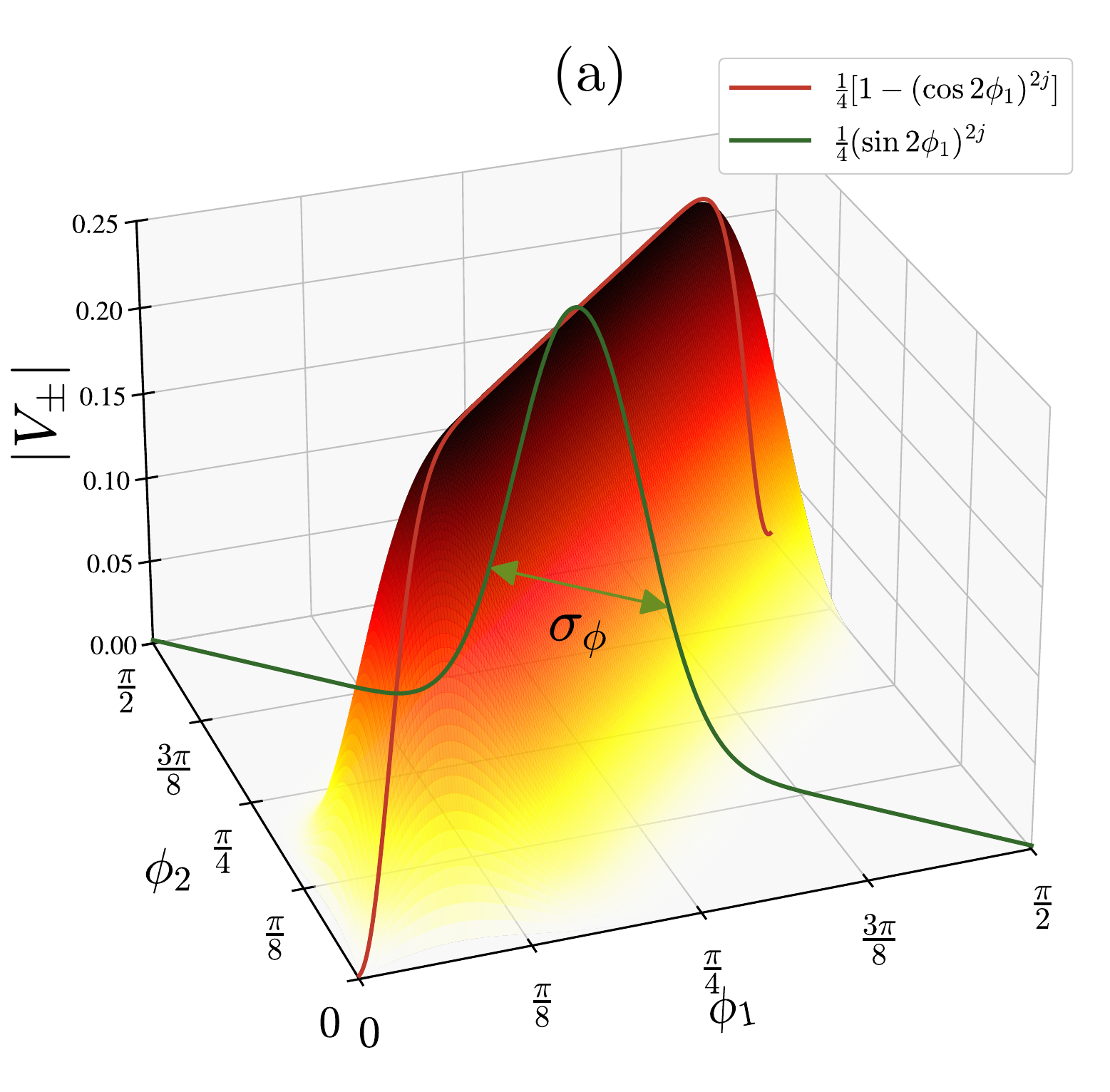}}
      \hfill
      \subfloat{\includegraphics[width=0.40\textwidth]{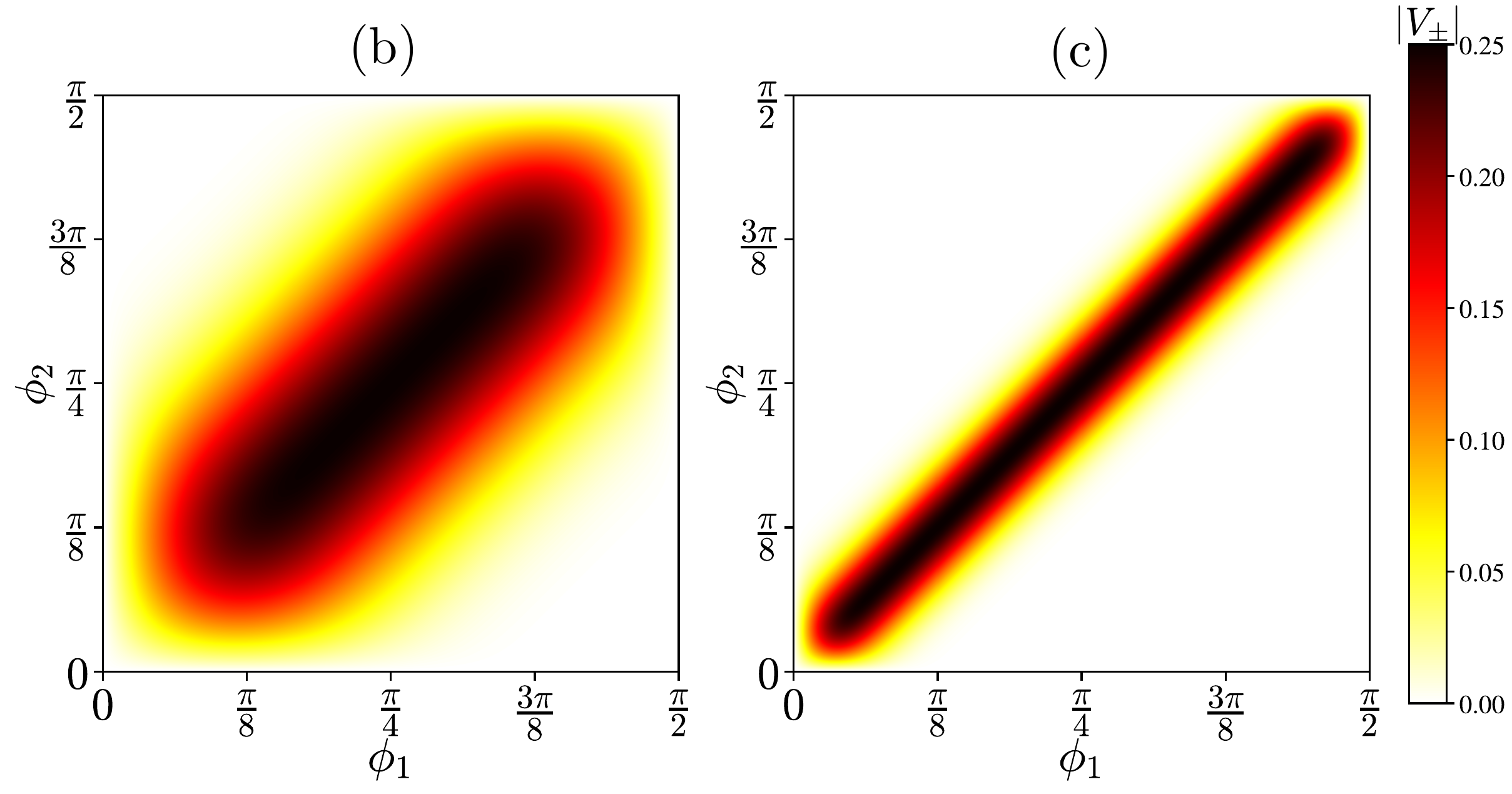}}
    \caption{Violation of NDC as function of two rotation angles $\phi_1$, $\phi_2$ for of (a) a spin $j=10$ with the analytically derived results for $\phi_1 = \phi_2$ being depicted in red and that for $\phi_1 = \pi/2 - \phi_2$ being presented in green, (b) for a spin $j=4$, and (c) for a spin $j=40$.}\label{fig:phi121}   
\end{figure}
Moreover, the range of $\phi_1$, $\phi_2$ with $\phi_1 \neq \phi_2$, for which $|V_{\pm}| \approx 1/4$, becomes \textit{narrower} with an increase in $j$. In order to quantify the precision required in the rotation angles, the case of $\phi_2 = \pi/2 - \phi_1$, which represents the largest possible rotation error, is studied in detail in Appendix~\ref{app:error_rotation}. 
Also in this case, an analytical solution is found, and the violation is given by,
\begin{equation}
\label{eq:violation_error_rot_an}
    \tilde{V}_{\pm}^{j=\text{even}} \left( \phi_1 \right) \approx \mp \frac{1}{4} \left(\sin 2 \phi_1 \right)^{2j} \;. 
\end{equation} 
For small errors of the form $\phi_1 = \frac{\pi}{4} + \delta \phi$, taking the Taylor expansion, 
\begin{equation}
    | \tilde{V}_{\pm}^{j=\text{even}} (\delta \phi)|  \approx \frac{1}{4} \left(1 - 2 \delta \phi^2 \right)^{2j} \propto \frac{1}{4} e^{- 4 j \delta \phi^2} \;.
\end{equation}
Let us assume that the error in the rotation angle $\delta \phi$ in different runs of the experiment is a random variable with normal distribution  $\mathcal{N} (0 , \sigma_\phi^2)$, where $\sigma_\phi$ is the standard deviation. Then it is possible to compute the violation as function of $\sigma_\phi^2$  by integrating $ | \tilde{V}_{\pm}^{j=\text{even}} (\sigma_\phi)| = | \int \text{d}(\delta \phi) \mathcal{N} (0 , \sigma_\phi^2)  \tilde{V}_{\pm}^{j=\text{even}} (\delta \phi)|$  (considering large enough number of experimental runs), such that
\begin{align}
    | \tilde{V}_{\pm}^{j=\text{even}} (\sigma_\phi)|  &= \frac{1}{4 \sqrt{1 + 8 j \sigma_\phi^2}} \;.
\end{align}
Thus, $| \tilde{V}_{\pm}^{j=\text{even}} (\sigma_\phi)| \approx 1/4$ for $\sigma_\phi \ll 1/( \sqrt{8 j})$ -- this is the bound for which such error is negligible. This represents one cause of quantum-to-classical transition: larger $j$ requires a higher rotation precision to detect NDC violation.

\subsection{Decoherence}

The study of decoherence becomes significantly important when considering macroscopic systems. The two main sources of decoherence in the proposed protocol are cavity decay and spin dephasing. Under this source of noise, the open dynamics is described by the Master Equation
\begin{align}
\label{eq:master}
    \frac{\partial \rho }{\partial t} &= i \left[ \rho, H_I \right] +\frac{\gamma_c}{2} \left(2 a \rho a^\dag  - \{ \rho(t) ,  a^\dag a \}   \right) \\
    &\hspace{1.9cm} + \frac{\gamma_s}{2} \left(2 J_z \rho J_z   - \{ \rho(t) ,  J_z^2 \}   \right)\nonumber \; , 
\end{align}
where $H_I$ is the Hamiltonian in Eq.~(\ref{eq:hamiltonian_off}), $\rho$ is the time-dependent density matrix of the cavity-spin joint system,  $\gamma_c$ and $\gamma_s$ are the decay and dephasing rates, respectively, and $\{ \cdot \; , \cdot \}$ represents the anti-commutator. This master equation is solved in Appendix~\ref{app:decoherence} and the density matrix at time $t$ is 
\begin{widetext}
\begin{align}
\label{eq:decoherence_dens}
    \rho(t)_{sc} &= \sum_{m,n =-j}^j d^{(j)}_{m, -j} \left(\phi \right) d^{(j)}_{n, -j}  \left(\phi \right) e^{i 2 \chi (m^2 - n^2) t } e^{-(m-n)^2 \gamma_s t } e^{-D_{m,n}( \gamma_c, t) }  \ket{j,m} \bra{j,n}_s \otimes \Ket{ \alpha_0 e^{ - \left(i m \chi + \gamma_c/2 \right) t}  } \Bra{ \alpha_0 e^{ - \left(i n \chi + \gamma_c/2 \right) t}  }_c \;, 
\end{align}
with
\begin{align}
\label{eq:decoherence_leaking}
      D_{m,n}(\gamma_c, t) &=\gamma_c |\alpha_0|^2 \left(\frac{1-e^{- \gamma_c t}}{\gamma_c } +\frac{e^{-t (\gamma_c +i \chi  (m-n))}-1}{\gamma_c + i \chi  (m-n)}\right) \;.
\end{align}
\end{widetext}

In order to keep the discussion general (without considering any specific physical system), we will compute the effect of the decoherence in terms of the ratios $r_c := \gamma_c/\chi$ and $r_s = \gamma_s/\chi$, i.e., in the unit of entanglement-coupling. See Appendix~\ref{app:decoherence}, for the details and for additional plots when both sources of decoherence are present simultaneously. We take the assumption that the decoherence effects are negligible during the rotations of the spin, compared to the entangling dynamics. This can be ensured as the rotation time $t_r$ can be made experimentally much shorter compared to the entanglement one $t_I$, by considering a high-intensity cavity rotation CS ($|\alpha_r\rangle_c$, as $\phi = 2 \alpha_r g t_{\text{R}} $). Furthermore, decoherence in the cavity will be dominated in the off-resonance interaction between the cavity and the spin, as a $2j+1$ dynamical superposition of cavity CSs entangled with the spin $m$-states will be created -- while in the resonance case, the cavity being in a single highly classical CS, the decoherence effect will be less prominent.

\begin{figure}[t!]
      \centering
      \subfloat{\includegraphics[width=0.49\textwidth]{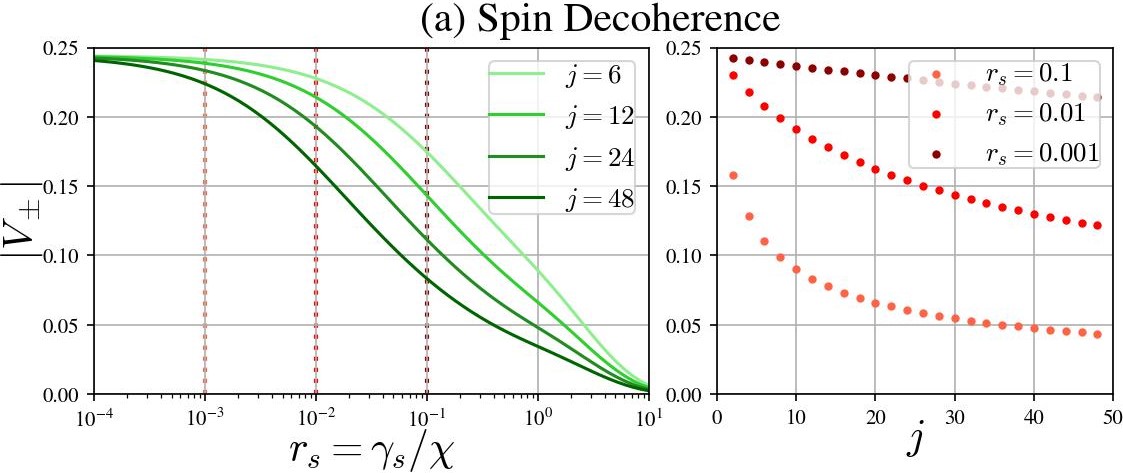}\label{plt:dec_spin}}
        \hfill
        \vspace{-0.2cm}
      \subfloat{\includegraphics[width=0.49\textwidth]{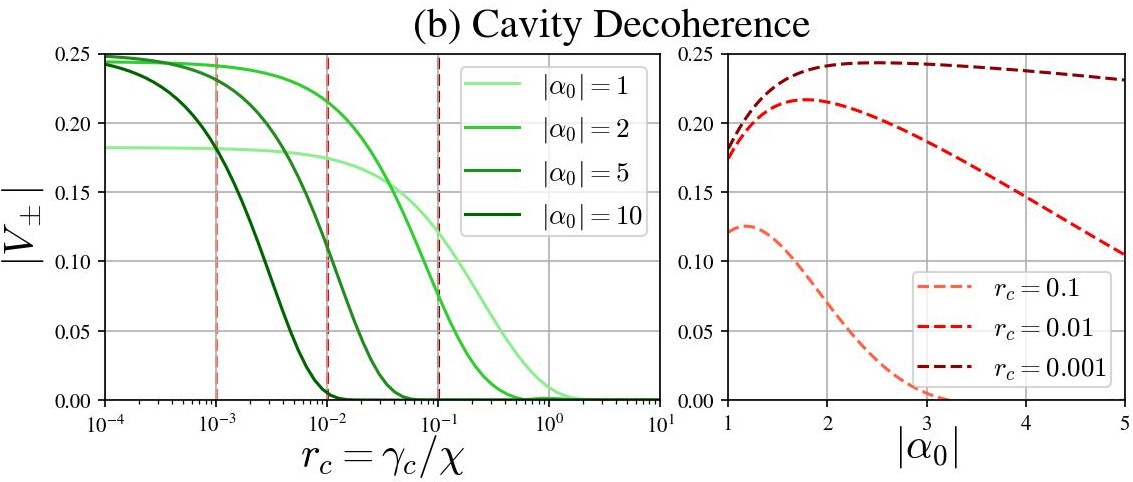}\label{plt:dec_cav}}
    \caption{Violation of the NDC of a large spin (a) under spin decoherence as function of the decoherence rate $\gamma_s$ for different spin values $j$, and (b) with cavity leaking, as function of the cavity decay rate $\gamma_c$ with different cavity intensity $|\alpha_0|$. The $\gamma_s$ and $\gamma_c$ are in the entanglement coupling ($\chi$) units.}
    \label{fig:dec_spin_caity}
\end{figure}

The violation of NDC at different cavity decay rates $r_c $ and at different spin decoherence rate $r_s$ -- considered independently -- is given in Fig.~\ref{fig:dec_spin_caity}. As the environmental noise increases (i.e. $r_c$ and $r_s$ increase), the violation is spoilt and the system starts behaving classically. 
In regard of the cavity decay, the effect of decoherence is found to be approximately constant among different spin values $j$, but sensitive to the choice of cavity field intensity $|\alpha_0|$ -- as shown in Fig.~\ref{fig:dec_spin_caity} and given by the $|\alpha_0|^2$ factor in Eq.~(\ref{eq:decoherence_leaking}). At time $t_I$, under decoherence, the cavity CS intensity is $\alpha_0(\gamma_c) = |\alpha_0| e^{- \frac{\pi}{2}\frac{\gamma_c}{ \chi} }$. When $\alpha_0(\gamma_c)$ is comparable to or smaller than one, which is the minimal uncertainty of the cavity CS, the measurement will not resolve the even/odd spin sub-spaces. This can be clearly seen in Fig.~\ref{fig:dec_spin_caity} for the case of $|\alpha_0| = 1$, in which -- even at very low decoherence rates -- $|V_\pm|$ does not approach the ideal value of $0.25$. However, it should be noted that as $r_c$ increases, lower intensity cavity CS are more resilient under cavity decoherence. Given a specific experimental setting (i.e. a value of $r_c$), an optimal cavity CS can be found such that $|V_\pm|$ will be maximum under cavity leaking. In what follows, the intensity of the cavity is kept at $|\alpha_0| = 2$, as it is resilient under cavity decoherence, but it still approaches the optimal value of $0.25$ as $r_c$ decreases (specifically, as $r_c \to 0$, $|V_\pm| \to 0.244$). The effect of spin decoherence becomes more prominent as the number of qubits increases. As shown in Fig.~\ref{fig:dec_spin_caity}, at a given $r_s$, the violation decreases as $j$ increases. Furthermore, as the environment becomes more noisy, the decay as function of the $j$ becomes faster. This source of decoherence can be mitigated by implementing well-known protocols (such as spin dephasing).

\subsection{Inhomogeneous Coupling}

In many experimental settings, inhomogeneity in the $N=2j$ qubit-cavity couplings can be present, that is, different qubits of the ensemble interact with different coupling strengths with the electromagnetic field. For large $N$, it is possible to assume that the probability distribution of the couplings approaches a Gaussian: $\{ g_i \sim \mathcal{N} \left(\braket{g}, \sigma_g^2 \right)\; \text{for} \; i \in [ 1, N]\}$, where $\braket{g} = g$ is the mean and $\sigma_g$ the standard deviation in the couplings. 

The exact evolution of each single qubit is computationally unreasonable, as the Hilbert space scales exponentially with the number of qubits. Hence, for simplifying calculations, the inhomogeneity in the couplings is studied by grouping together an equal number of qubits with similar couplings and assuming that all qubits in each sub-ensemble interact uniformly with the cavity. Mathematically, $N$ qubits are divided into $n$ subsets (ensuring that $N/n$ is integer). The average coupling of the $k$-th subset is given by
\begin{equation}
    \braket{g}_k = \frac{n}{N} \sum_{i = k N/n}^{(k + 1) N/n} g_{[i]},
\end{equation} 
where $k \in [1, n]$, and $[i]$ denotes the fact that the set of couplings $\{ g_i\}$ were first sorted in decreasing order (hence, similar qubits are grouped together). All qubits in the $k$-th subset ($\forall k \in [1,n]$) are assumed to interact uniformly with the cavity, with coupling $\braket{g}_k$. Hence, each sub-ensemble of qubits can be described by a single spin with total angular momentum $j_s =N/(2n)$ and Hilbert space $\mathcal{H}^{(k)} = C^{2j_s+1}$ ($\forall k \in [1, n]$). Assuming that the qubit-qubit interactions are negligible, the total Hilbert space $\mathcal{H} \sim \bigotimes_{k=1}^n \mathcal{H}^{(k)} $ for the whole duration of the experiment. Under this procedure, the cavity-spin ensemble interactions are given by: 
\begin{equation}
\label{inho_rot_ham}
    H_R  \approx  2 \alpha_r \sum_{k=1}^n \braket{g}_k J_y^{(k)}, \hspace{0.5cm} 
    H_I \approx \sum_{k=1}^n \braket{\chi}_k  J_z^{(k)} a^\dag a,
\end{equation}
where $J_m^{(k)}$ with $m= x,y,z$ are the spin operators of the $k$-th sub-spin, and $\braket{\chi}_k = \frac{4 \braket{g}_k^2}{\Delta}$. One should note that, as shown in Appendix~\ref{app:effective}, the parity projectors of Eq.~(\ref{projector_first_measurment}) commute with the $XY$-Hamiltonian ($H_{XY} = \frac{1}{4} \sum_{i\neq j}^N (  \sigma^{(i)}_x  \sigma^{(j)}_x + \sigma^{(i)}_y  \sigma^{(j)}_y )$). This implies that the additional qubit-qubit interactions in Eq.(\ref{eq:hamiltonian_off}), i.e., the $XY$-Hamiltonian, do not affect the protocol even at the single qubit-qubit case and hence can be omitted in the following discussion.

\begin{figure}[t!]
    \centering
    \includegraphics[width=0.49\textwidth]{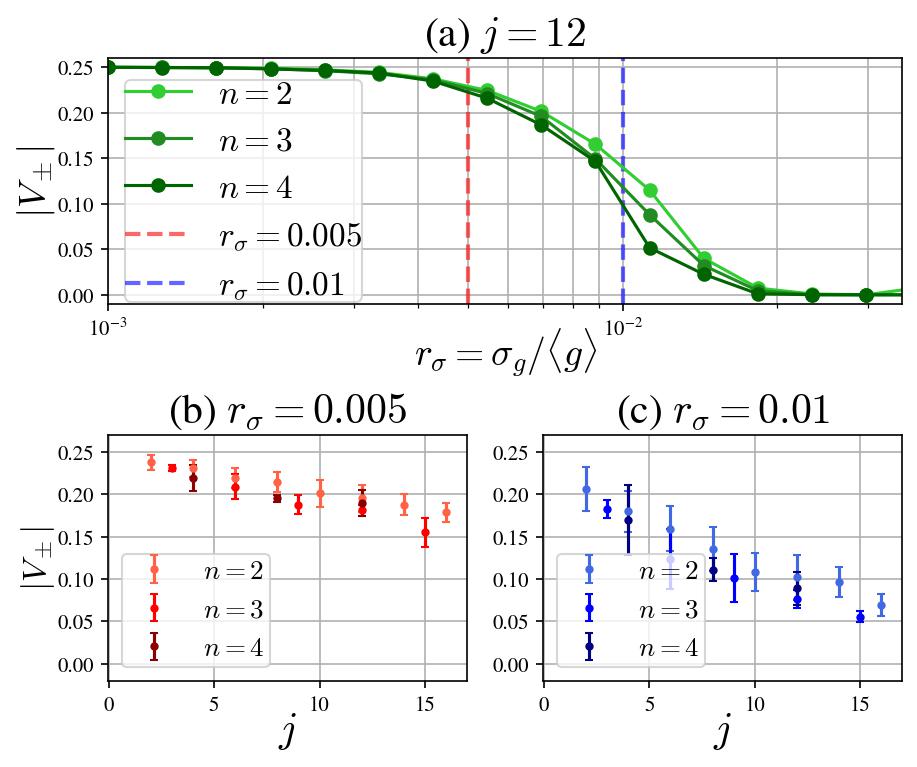}
    \caption{Violation of the NDC  of a qubit ensemble measured by a cavity field with inhomogeneous qubit-cavity couplings, extracted from a Gaussian distribution of standard deviation $\sigma_g$. Violation (a) as function of $r_\sigma =\sigma_g/\braket{g}$ with $j=12$; and as function of ensemble size at constant (b) $r_\sigma = 0.005$ and (c) $r_\sigma = 0.01$.}
    \label{fig:inoh}
\end{figure}

The same proposed protocol -- including the same interaction times such that $ t_I= \Delta \pi/(4 \braket{g}^2) $ -- with inohomoegenus couplings is discussed in details in Appendix~\ref{app:inohomgeneus} where the violation of NDC is derived. The numerical results as a function of the standard deviation $\sigma_g$ (in $\braket{g}$ units) are plotted in Fig.~\ref{fig:inoh}. 

As shown in Fig.~\ref{fig:inoh}, inhomogeneity in the couplings makes the system classical, and homogeneity is required to detect the quantumness of the ensemble. Larger ensembles require smaller inhomogeneity in the coupling, representing another form of quantum-to-classical transition. On the one hand, as $\sigma_g$ increases, the approximation to a single large spin becomes less accurate. On the other hand, a higher $n$ implies a more accurate description of an inhomogeneous qubit ensemble. The magnitude of $|V_\pm|$ decreases for larger values of both $\sigma_g$ and $n$. Quantumness of an ensemble $N \sim \mathcal{O}(10) $ can be found for $\sigma_g/\braket{g} \sim \mathcal{O}(10^{-2}) $. As $\sigma_g/\braket{g}$ decreases, the transition becomes less prominent with increasing ensemble size. 

We note that this effect also encodes the rotation uncertainty ($\sigma_\phi$), since the rotation error usually arises due to an inhomogeneity in the couplings between the resonator and the qubits.

\section{Quantum Measurement-Induced Collapse and Classical Disturbances\label{sec:classical_disturbance}}

Probing quantumness through the quantum violation of the NDC is primarily motivated by the fact that quantum measurement-induced collapse is an inherent part of quantum mechanical postulates. 
However, measurements of classical systems can also cause disturbances, specifically in experimental scenarios. Such disturbances, termed as ``Classical Disturbances'' (CDs), are not an inherent part of classical physics and, in principle, can be made arbitrarily small. This may not always be achievable in experiments, also known as \textit{clumsy-loophole} of the NDC inequalities. While various methods to reduce this loophole have been proposed, these are specific to the physical implementation and measurement procedure~\cite{knee2012violation,nsit3,largemass,gravity_measurement,loop1,loop2,loop3,loop4,loop5}. Here, we will describe two methods based on our parallel work \cite{LGIQC1} which are suitable for the protocol considered by us. 

The CD in the present case are caused by the \textit{difference} in the classical noise between the single and double measurement parts of the protocol\footnote{The single measurement part refers to the case when the final measurement is performed without the first measurement described in Sec. \ref{subsec:ideal}. On the other hand, the double measurement part is the case when the first and final measurements are performed.}. For instance, the difference between the experimental time durations in performing and not performing the first entangling dynamics induces a difference in the accumulated environmental noise, thus causing CD. The underlying idea to minimise the classical disturbance is to make the single measurement part of the experiment as \textit{similar} as possible to the double measurement part. 

One may ask if CDs, in fact, represent an experimental challenge and if the methods of CDs mitigation presented following are effective in decreasing such disturbances. In order to detect CDs, it is possible to note that for the rotation angle $\phi = \pi$ (or $\phi = 0$), the ideal quantum NDC violation is predicted to be zero. Thus, the detected violation of the NDC for such angles can be interpreted as being due solely to classical noise, assuming that the CDs are not function of the angle $\phi$.

In this section, two methods are presented to decrease the effect of CDs on NDC violation, in this context of this work. We are adapting these methods from the implementation of the protocol using IBM QCs (see our parallel work \cite{upcoming}), which highlighted the importance of avoiding CD to truly witness quantumness in experimental settings as there CDs were detected to cause approximately $30\%$ of the total ideal NDC violations (with $\phi = \pi$) and become negligible within statistical error under CDs reduction, thus making the protocol clumsiness-loophole free. Furthermore, as we shall see, the first method allows us to shed light on some subtle fundamental issues related to quantum measurement-induced collapse.



\subsection{Method 1 to reduce the classical disturbance} \label{method1}

In order to minimize the effect of the classical disturbance on the violation of NDC, we make the qubits-cavity entangling dynamics in the single and double measurement parts of the experiment the same. This can be achieved in the following steps (this is stated for integer $j$ -- the half-integer $j$ case will be qualitatively similar):

 \begin{itemize}
     \item In both parts of the protocol, after the initial rotation, the spin is entangled with the cavity such that the joint state becomes  $\ket{\phi_e}_{s} \Ket{ \alpha_0}_c + \ket{\phi_o}_{s} \Ket{- \alpha_0}_c$.

     \item Subsequently, the homodyne measurement is performed on the cavity, depending on which part of the protocol is being implemented -- the homodyne measurement is performed in the double measurement part. In contrast, it is not performed in the single measurement part.  On the one hand, if the measurement is performed, the spin-cavity collapses to one of the two unnormalized post-measurement states $\ket{\phi_e}_{s} \otimes \ket{\alpha_0}_c$ or $\ket{\phi_o}_{s} \otimes \ket{- \alpha_0}_c$ with the probabilities of Eq.~(\ref{eq:prob_1_ideal}). On the other hand, if the measurement was not performed, the state remains the same.

     \item In both cases, an identical spin-cavity entanglement dynamics is performed again. If the measurement was performed,  the second spin-cavity entanglement dynamics will evolve the aforementioned two collapsed states to $\ket{\phi_e}_{s} \otimes \ket{\alpha_0}_c$ and $\ket{\phi_o}_{s} \otimes \ket{\alpha_0}_c$ respectively.  If the measurement was not performed, the non-collapsed state under the second entangling dynamics evolves to the initial product state, that is, $\ket{\phi}_s \otimes \ket{\alpha_0}$. 

     \item In both of these cases, the cavity is then discarded trivially. Hence, the spin is in the $\ket{\phi_e}_{s}$ or $\ket{\phi_o}_{s}$ states if the homodyne measurement was performed, while in the $\ket{\phi}_s$ state if the homodyne measurement  was not performed. 
     
     \item   The second rotation of the spin and the final measurement (involving another spin-cavity entanglement dynamics along with the homodyne measurement on the cavity) are performed in both cases.

 \end{itemize}
 
In this modified protocol, the entangling dynamics is performed three times in both the double-measurement and single-measurement cases. The only difference between the two parts of the protocol is the intermediate homodyne measurement on the cavity. The violation of NDC is equivalent to that of the ideal protocol depicted in Sec.~\ref{sec:ideal_case}, but ensuring that the classical disturbance is minimised.

\textit{Note Added:} The above-modified protocol points out a fundamental feature of the quantum measurement process. Quantum measurement consists of two stages: (Step 1:) entangling dynamics between the target system (the system to be measured -- the qubit-ensemble in the present case) and the probe (cavity in the present case), (Step 2:) reading out the outcome by performing the projective measurement on the probe (homodyne measurement on the cavity in the present case). The Step 1 causes disturbance on the target system, as any pure state of the target system becomes mixed after the entangling dynamics. On the other hand, the Step 2 makes this disturbance irreversible, as the pure entangled joint state of the target-probe collapses into a specific state depending on the outcome, thus becomes mixed due to this step over many runs of the experiment\footnote{In a many-world interpretation, when reading out the outcome of a measurement, the target-probe entangles with the rest of the world, under a unitary that cannot be reversed.}. In the absence of this Step 2, the disturbance on the target can be eliminated by applying the inverse of the entangling unitary dynamics on the joint state of the target-probe (this happens in the above-mentioned modified protocol in the absence of the homodyne measurement). Hence, while the entangling dynamic is responsible for quantum measurement-induced disturbance, the reading-out step is crucial for making this disturbance irreversible. These two steps together cause an irreversible collapse on the target system -- a genuine non-classical feature.

Note that, without the reading out part, doing and then undoing the entangling dynamics between the probe and the system is done in the context Stern-Gerlach interferometry~\cite{gravity_measurement,SGI2,SGI1}.

\subsection{Method 2 to reduce the classical disturbance}\label{method2}

The aforementioned method requires three entangling dynamics, which may be experimentally difficult to realise. An alternative method to minimise the CD is to always perform two entangling dynamics, but change the input state of the probe between the single and double measurement parts of the protocol. 

The double measurement part of the protocol is the same as presented in Sec. \ref{sec:ideal_case}.  On the other hand, the single measurement part is the same as the double measurement part, except that the input state of the cavity during the first entangling dynamics is $(\ket{\alpha_0} + \ket{- \alpha_0})/\sqrt{2}$. Hence, after the initial rotation of the spin state and introducing the cavity, the spin-cavity state becomes $\ket{\phi}_s \otimes (\ket{\alpha_0}_c + \ket{- \alpha_0}_c)/\sqrt{2}$, which evolves under the entangling dynamics to $\ket{\phi}_s \otimes \ket{\alpha_0}_c$. Subsequently, the homodyne measurement on the cavity is performed trivially, leading to the post-measurement state of the spin $\ket{\phi}_s$. The second rotation and measurement are then performed as usual, leading to the same measurement probabilities, and the same NDC violation of Sec. \ref{sec:ideal_case}. 

In this method, the first entangling dynamics in the single measurement part does not entangle the spin and the cavity (we are using the term ``entangling dynamics" just to refer to the interaction between the spin and the cavity mentioned in stage (iii) of Sect. \ref{subsec:ideal}). Consequently, this interaction along with the homodyne measurement on the cavity cannot be interpreted as a measurement of the spin at all. Hence, in effect, in the single measurement part, measurement on the spin is performed only once -- i.e., the final measurement on the spin. Thus, the only difference between the single and double measurement parts of this method is applying an (optional) local unitary on the probe, significantly reducing the CD. It should be noted that such states are routinely prepared in optical systems~\cite{cat1,cat2}. 

After adopting either of the two aforementioned methods for reducing CD, a control experiment should be performed in which the ideal quantum violation is predicted to be zero (for example, for $\phi=0,\pi$). Any observed NDC violation in this control experiment can then be attributed solely to CD. The experimentally observed NDC violation for other values of $\phi$ can be considered as a conclusive signature of quantumness (i.e., ruling out any macrorealist description) only if the difference between the observed NDC violation and the NDC violation arising solely from CD in the control experiment is significant within the statistical uncertainties~\cite{LGIQC1,loop2}).

\section{Discussion: Physical Implementation \label{sec:physical_system}}

\begin{figure}[t!]
    \centering
    \includegraphics[width=0.45\textwidth]{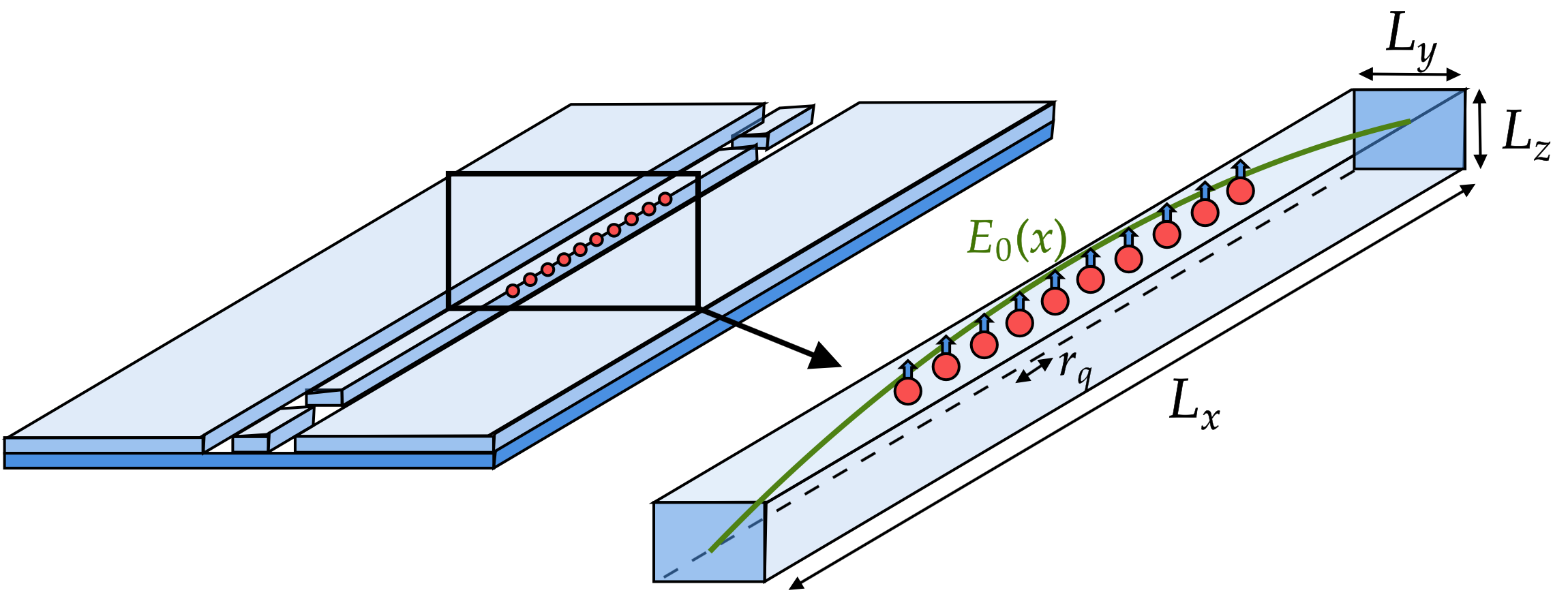}
    \caption{Diagram representing the chosen configurations of the qubit-ensemble placed at distance $r_q$ electrically coupled to the fundamental electric mode $E_0(x)$ of a CWR.}
    \label{fig:diagram_exp_SetUp}
\end{figure}

In numerous physical systems, the control and measurement of qubits is implemented through a cavity field or resonator, representing the potentiality and flexibility of the presented protocol, which this section explores. In the following, we investigate the parameter spaces -- for instance, frequencies and couplings --  of current experiments in relation to the noise analysis performed in Sec.~\ref{sec:quantum_classical}. The parameters considered are summarised in Table \ref{tab:parameters}, where the maximum number of qubits for which NDC can be violated ($N_{\text{NDC}}$) is given for different physical systems, representing one of the main results of the work. Here, we give a comparitive analysis between (a)~superconducting qubits, (b)~Rygberg atoms, and (c)~spin qubits with artificial spin-orbit interactions (e.g. double quantum dots or  dopant hole spins), coupling a superconducting Coplanar Waveguide Resonator (CWR), experimentally achieved in numerous systems~\cite{Xiang2013}. In Fig. \ref{fig:diagram_exp_SetUp}, a schematic representation of the experimental setup is given.

The experimental parameters defining the physical system are the qubit frequency $\omega$, the average single qubit-cavity coupling $g$, and the detuning frequency $\Delta$. The experimental noises are captured by: (a) the rotation uncertainty $\sigma_\phi$; (b) the cavity decay rate $\gamma_c/(2 \pi)$, (c) the spin dephasing rate $\gamma_s/(2 \pi) =  T_2^{-1}$ (with $T_2$ being the dephasing time of each qubit); and (d) the inhomogeneity in the couplings $\sigma_g$. \textit{Given such parameters and a physical implementation of the experiment, what is the maximum achievable number of qubits $N_{\text{NDC}}$ such that the violation of NDC is detectable?} 

Before investigating the specific implementation (and the constraints that it implies), general considerations can be drawn from the noise analysis already performed. First of all, we define \textit{acceptable} noise, as the amount of noise for which the predicted NDC signal for $\phi = \pi/4$ is $\sim 0.1$. This requires $\sim 10^4$ experimental runs to ensure that the statistical error is one order of magnitude lower than the
detected NDC violations (since $n$ number of runs
can determine the probability of the outcomes with uncertainty $\sim 1/\sqrt{n}$). Thus, one finds that the noise effects are acceptable: (a) for cavity decay --  if $\gamma_c  \lesssim \mathcal{O} (0.1 \chi)$, by choosing the optimal cavity strength ($|\alpha_0|$) such that the NDC violation is maximised given the aforementioned $\gamma_c$ (see Fig. \ref{fig:dec_spin_caity}); (b) for spin decoherence -- if $\gamma_s  \lesssim \mathcal{O} (\chi/ N)$ (see Fig. \ref{fig:dec_spin_caity}), (c) for coupling inhomogeneity -- if $\sigma_g  \lesssim \mathcal{O} (g /(10 N))$ (see Fig. \ref{fig:inoh}), and (d) for rotation precision -- if $\sigma_\phi  \lesssim \mathcal{O} (1/ (2 \sqrt{ N}))$. 

We note that both bounds on the decoherence rates scale linearly with the entanglement coupling $\chi = 4 g^2/ \Delta$, which should therefore be maximised. This is ensured by decreasing $\Delta$ and increasing $g$. Given that the off resonance regime must hold ($\omega_0 \gg \Delta \gg g$), the minimum value of $\Delta$ is limited, and we can take $\Delta \sim 10 g$ as an appropriate choice, for all implementations~\cite{PhysSuper1,PhysChoice,Pritchard2014}. 
This implies $\chi \sim 0.4 g$. Inverting the bounds for the decoherence noise, we find that for $g \gtrsim \mathcal{O} (25 \gamma_c )$ and $g \gtrsim \mathcal{O} ( 2.5 N \gamma_s)$,  the decoherence effects are acceptable. 
In order to detect violation of NDC, the single qubit-photon coupling must be greater than the cavity decay rate and spin dephasing rate, i.e., \textit{the strong coupling regime is a requirement of the experiment}, which has been experimentally achieved in numerous architectures, with also multiple qubits~\cite{PhysSuper2,PhysSuper3,PhysSolid1,PhysSolid2,PhysSolid3,Phys2Super1,Phys3Super1}. For this reason, we restrict ourselves to the case where the qubit and the resonator couple via the electric field, and exclude from the analysis  magnetic coupling (as it is found to be too weak for the protocol's realisation, where only effective multi-qubit couplings allow strong coupling).

\begin{table*}[t]
\renewcommand{\arraystretch}{1.3}
\begin{tabular}{ |p{5cm}|p{1.2cm}|p{1.0cm}|p{2.2cm}|p{2.2cm}|p{2.2cm}|  }
        \hline
        Experimental Parameters & Symbols  & Units & 
        Superconducting Qubits &  
         Rydberg Atoms &  
        Spin Qubits with Spin-Orbit\\
         \hline
        \hline
        Electric Dipole & $d$ & $a_0 e$ & $n_d \sim 10^4$ & $ n_d\sim 3 \cdot 10^3 $ & 
        $n_d \sim 10^2 $ \\
        \hline
        Single Qubit-Photon Coupling & $g/(2\pi)$  &  MHz & $ 110$ &   $ 7$ &   $ 1$  \\
        Detuning Frequency & $\Delta/(2\pi)$ & MHz & $ 1 000$  & $ 70$ &  $ 10$ \\
        Entanglement Coupling &  $\chi/(2\pi)$ &  MHz& $ 45$   &  $ 3$   & $ 0.4$    \\
        Distance Between Qubits &  $r_q$ &  $\mu$m & $ 35$    &  $ 23$   &  $ 7$    \\
        \hline
        \hline
        \multicolumn{6}{|c|}{Bound on Number of Qubits and Decoherence Rates} \\
        \hline
        \hline
     Bound on number from inhomogeneity & $N_{\text{NDC}}^{\text{inh}}$  & NA   &$\lesssim 41$    &$\lesssim 53$  & $\lesssim 110$    \\
        Spin Dephasing Rate & $\gamma_s/(2\pi)$  & kHz   & $ \lesssim4500$  & $ \lesssim300$  &   $\lesssim 45$  \\
        Cavity Decay Rate & $\gamma_c/(2\pi)$  & kHz   & $\lesssim 1000$  & $\lesssim 54$  &   $\lesssim 24$  \\
        Single-qubit Rotation Uncertainty & $\sigma_\phi$  & rad  & $\lesssim 0.078$  & $\lesssim 0.068$  &   $\lesssim 0.047$  \\
        \hline
    \end{tabular}
  \caption{Orders of magnitude of the parameters for realising the NDC parity protocol with (i)~Superconducting Qubits, (ii)~Rydberg Atoms, and (iii)~Spin qubits. For all of the above, the fundamental frequency of the CWR is $\omega/(2 \pi) \approx 3.5$ GHz ($e$~is the electron charge, and $a_0$ the Bohr radius).}
  \label{tab:parameters}
\end{table*} 

In Appendix~\ref{app:physical}, we discuss the electromagnetic field generated by cavities of the form given in Fig.~\ref{fig:diagram_exp_SetUp}, i.e. a superconducting box without one side parallel to the $xy$ plane, at hight $L_z$, which is used to approximate the CWR. The classes of microwave modes generated by the boundary conditions imposed by the box are the Transverse Magnetic or Electric Modes, where the former, here chosen for the coupling with the qubits, has a vanishing magnetic field component in the propagating $z$-direction~\cite{transmission_lines,Pritchard2014}. Keeping in mind fabrication constraints~\cite{CWR} (and Fig.~\ref{fig:diagram_exp_SetUp}), the dimensions of the CWR chosen are: for the long side, $L_x \sim 2 \cdot 10^{-2}$ m, for the short side, $L_y \sim 8 \cdot 10^{-6}$ m, and for the thickness, $L_z \sim 2 \cdot 10^{-7}$, such that the volume is $V \sim 3.2 \cdot 10^{-14}\text{ m}^3$. 

The presence of the substrates at $z<0$, induces an effective permittivity $\epsilon_r \sim 5$, which gives the resonance fundamental frequency of $\omega/(2 \pi) \sim c/(2 L_x \sqrt{\epsilon_r})\sim 3.5$ GHz ~\cite{CWR}.  Furthermore, the absence of the closing side of the resonator at $z = L_Z$ implies an exponential decay of the electric field in the $z$ direction, for which the modes are said to be evanescent~\cite{transmission_lines}. As shown in the Appendix~\ref{app:physical}, this effect is negligible within the resonator, such as in the case of the superconducting and spin qubits, while when the qubits are placed outside the resonator, as in the case of Rydberg atoms (usually trapped by optical tweezers, at $z\approx 50 \; \mu$m~\cite{Pritchard2014,PhysRygberHogan}), the decay strength of the electric field $z$-component is decreased by a factor of $ \sim 0.2$. We shall consider this specific effect in Table~\ref{tab:parameters} for the case of Rydberg atoms (which decreases $g$ by one order of magnitude), while we will omit the explicit $z$-dependency in the following expressions.

Let us consider a qubit with dipole $\mathbf{d} = (0,0,d)$ placed in the centre of the cavity, i.e. $(x,y) = (0,0)$, interacting with the fundamental mode of the CWR. The dipole can be written as $d = n_d a_0 e$, where $a_0$ is the Bhor radius, $e$ is the electron charge, and $n_d$ characterises the type of qubit (see Table~\ref{tab:parameters}). Thus, the qubit-resonator coupling is given 
 \begin{equation}
 \label{eq:physical_coupling}
    g = \frac{\mathbf{E}_0 \cdot \mathbf{d}}{\hbar} \sim d \sqrt{ \frac{\omega}{2 \hbar \epsilon_0 \epsilon_r  V} } \sim  2 \pi  \cdot n_d  \cdot 71 \; \text{kHz} \;,
\end{equation} 
where $E_0$ is the vacuum electric field density, $V$ is the volume of the resonator, and $\epsilon_0$ and $\epsilon_r$ are the vacuum and relative (effective) permittivities. From the values of $g$ given in Table~\ref{tab:parameters}, it is possible to note that the bound on decay rate of the resonator does not represent an experimental challenge, as lower decay rates have been reported in a large number of resonator, for instance, $\sim 1$ kHz~\cite{guide1}. Here, it should be noted that a higher decay rate may be preferable in order to collect the necessary light to perform homodyne detection~\cite{delva2024}.

The number of qubits achievable ($N_{\text{NDC}}$) is restricted by the volume in which the qubits can be placed. On the one hand, a large qubit-cavity coupling $g$ -- which requires a small volume -- is preferable such that the experiment is performed within the coherence time of the system and the strong coupling regime is met. On the other hand, a small volume increases the noise caused by direct qubit-qubit interactions -- as they are placed closer to each other -- and by coupling inhomogeneity -- as the electric field strength of the resonator would change more rapidly among qubits, implying different couplings. 
The dipole-dipole interaction (with coupling $d_g$) is negligible compared to the entanglement interaction (with coupling $\chi \sim 0.4 g$, which is the slowest process of the experiment), when
\begin{equation}
    d_g \sim   \frac{d^2 }{4 \pi \epsilon_0 \hbar r_q^3} \ll  \chi \implies r_q \sim 2 \left(\frac{d^2 }{ \hbar \epsilon_0 g} \right)^{1/3} .
\end{equation}
This gives the minimum distance between qubits ($r_q$) at which they can be placed such that the dipole-dipole interaction is negligible. Choosing $d_g/\chi \sim 0.05$, the minimum distance is given by $r_q \approx 1.6 \cdot n_d^{1/3} \cdot \mu\text{m}$.

However, different positions of the qubits in the resonator introduce inhomogeneity in the coupling because different qubits experience a different electric field strength due to its dependency on $x$. Specifically, the qubits are placed in a line at $(x,y) = (k r_q, 0)$, where $k \in [-N/2, N/2]$ and the $z$-component of the electric field density is the first harmonic $E_z(x) \sim E_0 \cos(\pi x/L_x)$. Then, the coupling of the $k$-th qubit is given by
\begin{equation}
\label{eq:inho_phyiscal}
     g_{k} = g \cos \left(  \frac{\pi r_q}{L} k \right)  \;.
\end{equation}
Such an inhomogeneity increases with the number of qubits as $k \in [-N/2, N/2]$ and it is inversely proportional to the volume ($V \sim L$). In Appendix \ref{app:physical}, the arising inhomogeneity is studied and found to be negligible (i.e. $\sigma_g/g  \lesssim \mathcal{O} (1/(10 N))$, as shown in Sec.~\ref{sec:quantum_classical}), if the following bound is met
\begin{equation}
    N_{\text{NDC}}^{\text{inh}}\lesssim 0.6 \left( \frac{L_x}{r_q} \right)^{2/3} \sim  \frac{320}{n_d^{2/9}}\;.
\end{equation}
One should note that this may not be a stringent bound compared to one given by physical implementation where fabrication imperfections (in the case of superconducting and spins qubits) and uncertainty in the trapping positions (Rydberg atoms) may increase inhomogeneity effects. However, the source of inhomogeneity studied may be decreased via the use of multiple cascaded cavities, similar to~\cite{Ruskov2003,Lalumiere2010,Roch2014,PhysSolidMeasureBased,delva2024}.

The chosen parameters are given in Table~\ref{tab:parameters}, for which the bound given by the inhomogeneity ($N_{\text{NDC}}^{\text{inh}}$) is derived and subsequently used to compute the bounds on the spin dephasing rate and single-qubit rotation uncertainty, which have a $N$ dependency -- i.e. $\gamma_s \lesssim g /( 2.5 N_{\text{NDC}}^{\text{inh}})$ and  $\sigma_\phi  \lesssim \mathcal{O} (1/ (2 \sqrt{ N_{\text{NDC}}^{\text{inh}}}))$ . On one hand, it is possible to note that for highly interacting qubits, as in the case of superconducting qubits and Rydberg atoms with high dipoles (and hence coupling), the dephasing bound is not stringent, as the experiment can be performed within decoherence times. However, high coupling increases the direct dipole-dipole interactions, which subsequently increases the minimum distance at which the qubit can be place, and hence the inohomodeinty bound limits the experiment to $N_{\text{NDC}}^{\text{Sup}} \sim 41 $ and $N_{\text{NDC}}^{\text{Ryd}} \sim 53$, for superconducting qubits and Rydberg atoms, respectively. One the other hand, in the case of semiconductor spins, the lower dipole increases the bound given by the inhomogeneity $N_{\text{NDC}}^{\text{Spin}} \sim 110 $, while, however, the lower coupling increases the demand on coherence time. 

\section{Conclusions \label{sec:conclusion}}

We present a protocol to implement parity measurements on an ensemble of qubits exploiting their dispersive interaction with a resonator and a subsequent homodyne measurement of a resonator's quadrature. By comparing  single and double parity measurement schemes, a constant quantum violation of MR is detectable via NDC with arbitrary many qubits: in the ideal case, the non-classicality of a qubit ensemble is always detectable, even for many $\hbar$ units. The quantumness manifests itself via the unavoidable disturbance caused by collapse of the ensemble's wavefunction under a parity measurement. The proposed protocol is operationally \textit{independent} of the size of the ensemble, 
and provides an experimentally implementable parity measurement-based scheme to test the non-classicality of macroscopic qubit ensembles.

We find that the limiting factors of the proposal are induced by decoherence and inhomogeneity in the qubit-resonator couplings, which induces the quantum-to-classical transition, recovering the classical behaviour of the macroscopic world as the number of qubits increases. This behaviour is not fundamental; rather it is only due to operational constraints, as it becomes more challenging to perform the experiment with a larger number of qubits.

We detail a specific implementation of the protocol using CWR with a variety of qubits, namely, superconducting qubits, Rydberg atoms, and spin qubits with artificial spin-orbit interaction. In this context, it should be noted that operationally similar experiments -- i.e. dispersive interaction and homodyne measurement via CWR --  have already been performed for probabilistic generation of the entangled state between two qubits (effectively performing a parity measurement), for superconducting and spin qubits~\cite{Roch2014,PhysSolidMeasureBased,delva2024}. Thus, our generalizations to $N$-qubits is already accessible for experimental implementations to test the macroscopic limit of quantum mechanics.

We conclude that, given the state-of-the-art quantum technologies, it is possible to detect the quantumness of ensembles of superconducting qubits, Rydberg atoms, and spins in semiconductors with spin-orbit interaction up to $N_{\text{NDC}}^{\text{Sup}} \sim 41 $, $N_{\text{NDC}}^{\text{Ryd}} \sim 53 $, and $N_{\text{NDC}}^{\text{Spin}} \sim 110 $, respectively. The proposed protocol is scalable in the number of qubits as the quantum technologies progress to more isolated systems, such that the transition to classicality can be experimentally investigated under different noises. Here we note our parallel work in the context of QC~\cite{upcoming}, where the clumsiness-loophole-free protocol has been used to detect quantumness of IBM QCs up to $38$ qubits, showing the transition to classicality as the QC becomes macroscopic.  

 We emphasize that the use of collective parity measurement as the intermediate measurement is a key technical ingredient of our proposal. While parity measurements are well established in quantum information processing and quantum metrology~\cite{Parity1,Parity2,Parity3,Parity4}, their role here is to provide a practical tool for experimentally probing the nonclassicality of a macroscopic qubit-ensemble.

To conclude, the presented protocol explores the quantum-to-classical transition in macroscopic qubit ensembles. The results provides an explicit demonstration of the fact that the emergence of classical behaviour in the large-spin limit is closely connected to practical limitations in controlling imperfections and decoherence. Our work presents a suitable protocol, realizable with a variety of physical systems, to demonstrate nonclassical features even for systems comprising a large number of quantum constituents using current technologies. 


\section{Acknowledgments}

We thank Farhan Hanif for his contributions at the initial stages of this work. The contributinos of BZ and LB were supported by the UK Engineering and Physical Sciences Research Council (EPSRC) through grants EP/R513143/1 and EP/T517793/1; and EP/R513143/1 and EP/W524335/1, respectively. DD is grateful to the Royal Society, United Kingdom, for support through a Newton International Fellowship no. NIF$\backslash$R1$\backslash$212007. DD and SB acknowledge financial support from the EPSRC through grants EP/X009467/1 and EP/R029075/1, and the UK Science and Technology Facilities Council (STFC) through grants ST/W006227/1 and ST/Z510385/1. SDH acknowledges support from the EPSRC through grant grant EP/Y022688/1. DD and SB also acknowledge the financial support from the Royal Society, UK under the scheme ``Newton International Fellowships
Alumni 2025'' (Grant No. AL$\backslash$251043). DD gratefully acknowledges University College London for its kind hospitality during his visit in July 2026, which provided the opportunity to complete a substantial part of this work.

\onecolumngrid

\appendix

\section{Off-Resonance Effective Hamiltonian\label{app:effective}}

In the rotating wave approximation and interaction picture, the interaction Hamiltonian between $N$ qubits and a common resonator -- with de tuning frequency $\Delta$ -- is given by $H_{\text{JC}}^{\Delta} (t) = i \sum_{i=1}^N  g_i (e^{i \Delta t} a \sigma_+^{(i)} + e^{- i \Delta t}  a^\dag \sigma_-^{(i)}) $. In the limit of $g \ll \Delta $, the first order term in the Dyson series is negligible (as $\sim \mathcal{O} (g/\Delta)$), such that the leading order of the unitary is given by the second one 
\begin{align*}
    U_{I}(t) &\approx \mathds{1} - \int_{0}^t \text{d}t'  H_{\text{JC}}^{\Delta} (t') \int_{0}^{t'}\text{d}t'' H_{\text{JC}}^{\Delta} (t'') \\
    &\approx \mathds{1} -\sum_{i}^{N}  \frac{g_i^2}{\Delta} \int_{0}^t \text{d}t'  \left(e^{i \Delta t'} a \sigma_+^{(i)} + e^{- i \Delta t'}  a^\dag \sigma_-^{(i)}\right)  \left(\left( e^{i \Delta t'} - 1 \right) a \sigma_+^{(i)} - \left( e^{-i \Delta t'} - 1 \right)  a^\dag \sigma_-^{(i)}\right)  \\
    & \hspace{1cm} - \sum_{i\neq j}^{N}  \frac{g_i g_j}{\Delta} \int_{0}^t \text{d}t' \int_{0}^{t'}\text{d}t''  \left(e^{i \Delta t'} a \sigma_+^{(i)} + e^{- i \Delta t'}  a^\dag \sigma_-^{(i)}\right)  \left(\left( e^{i \Delta t'} - 1 \right) a \sigma_+^{(j)} - \left( e^{-i \Delta t'} - 1 \right)  a^\dag \sigma_-^{(j)}\right)  \\
    &\approx \mathds{1} - i t \left[ \sum_{i=1}^N \frac{2 g_i^2 }{ \Delta } a^{\dag} a \sigma_z^{(i)} + \sum_{i\neq j}^N \frac{g_i g_j}{ \Delta }  \left( \sigma^{(i)}_+  \sigma^{(j)}_- +  \sigma^{(i)}_-  \sigma^{(j)}_+ \right)   \right] = \mathds{1} - i t \left[ \sum_{i=1}^N \frac{2 g_i^2 }{ \Delta } a^{\dag} a \sigma_z^{(i)} + \sum_{i\neq j}^N \frac{g_i g_j}{ \Delta }  \left( \sigma^{(i)}_x  \sigma^{(j)}_x +  \sigma^{(i)}_y  \sigma^{(j)}_y \right)   \right]
\end{align*}
where the third equality follows from $[ a, a^\dag] = \mathds{1}$,$[ \sigma_+, \sigma_-] = \sigma_z$, and neglecting the $\mathcal{O} (g^2/\Delta^2)$ terms.  This implies that the effective Hamiltonian is given by the first line of Eq.~(\ref{eq:hamiltonian_off}). Using the definition of collective operator $J_{\alpha} = \frac{1}{2} \sum_{i}^{N} \sigma_{\alpha}^{(i)}$  and the fact that the total angular momentum is given by
\begin{align*}
    \mathbf{J}^2  &= J_x J_x + J_y J_y + J_z J_z = \frac{1}{4} \sum_{i,j}^N \left(  \sigma^{(i)}_x  \sigma^{(j)}_x + \sigma^{(i)}_y  \sigma^{(j)}_y\right) + J_z J_z =\frac{1}{2} \mathds{1} +  J_z J_z  +   \frac{1}{4} \sum_{i\neq j}^N \left(  \sigma^{(i)}_x  \sigma^{(j)}_x + \sigma^{(i)}_y  \sigma^{(j)}_y \right)
\end{align*}
where the second line follows from the fact that $\sigma_j^{(i)}\sigma_j^{(i)} = \mathds{1}$, for $j\in \{ x,y,z \}$ and we recognize the qubit-qubit term in the Hamiltonian. Inverting the expression and assuming $g_i = g \; \forall \; i$, and substituting in the effective Hamiltonian to get Eq.~(\ref{eq:hamiltonian_off}). 

One may further note that this additional qubit-qubit term ($H_{XY} = \frac{1}{4} \sum_{i\neq j}^N \left(  \sigma^{(i)}_x  \sigma^{(j)}_x + \sigma^{(i)}_y  \sigma^{(j)}_y \right)$) does not affect the protocol even at the qubits level (and not at the large spin), which shall become useful during noise analysis. Let the complete computational basis be $\ket{J} = \bigotimes_{i = 1}^{N} \ket{j_i}$ of the Hilbert space of the $N$-qubits, i.e. $\mathcal{H} \sim \left( \mathds{C}^{2} \right)^{\otimes N}$, with $j_i = \{0, 1\}$ and $J$ being the string $\{j_1, ..., j_N\}$, labeling a state. One may note that 
\begin{equation}
    H_{XY} \ket{J} = \sum_{i\neq k}^N \frac{g_i g_k}{ \Delta }  \left( \sigma^{(i)}_x  \sigma^{(k)}_x +  \sigma^{(i)}_y  \sigma^{(k)}_y \right) \ket{j_1... j_i ... j_k .... j_n}
    = \sum_{i\neq k}^N \frac{g_i g_k}{ \Delta }  \ket{j_1 ... j_k, ... j_i ... j_n}
\end{equation}
which can be used to prove that observables $\Pi_{\pm}$ commutes with the Hamiltonian. In fact, for instance, the even subspace is spanned by the vectors $\{ \ket{J} \}$ with $J \in \mathcal{S}_{\text{e}}$, where $\mathcal{S}_{\text{e}}$ is the set of strings such that $\sum_{i = 1}^N j_i$ is even. In this form, the POVM given in Eq.~(\ref{projector_first_measurment}), can be written as $\Pi_{+} = \sum_{ J \in  \mathcal{S}_{\text{e}} } \ket{J} \bra{J}$. Then, 
\begin{equation}
    \left[ \Pi_+ , H_{XY} \right] =  \sum_{i\neq k}^N \frac{g_i g_k}{ \Delta }  \sum_{ J \in  \mathcal{S}_{\text{e}} }  \left( \ket{j_1 ... j_k ... j_i ... j_n} \bra{j_1 ... j_i ... j_k ... j_n} - \ket{j_1 ... j_i ... j_k ... j_n} \bra{j_1 ... j_k ... j_i ... j_n}  \right) 
\end{equation}
Give that the first summation is over all the strings with $\sum_{i = 1}^N j_i$, which is independent on the ordering of the eigenvalues, the commutator is trivially zero as the $H_{XY}$ Hamiltonian only swaps the order. 

Similarly, it is the case for the $\hat{J}_z$ operator. At the level of the qubits: 
\begin{align}
    [J_z, H_{XY}] &=  \left[ \sum_{i} \sigma_z^{(i)}, \sum_{jk} \frac{g_j g_k}{\Delta} \left(\sigma_{x}^{(j)} \sigma_{x}^{(k)} + \sigma_{y}^{(j)} \sigma_{y}^{(k)} \right) \right] \nonumber \\
    &= \sum_{jk} \frac{g_j g_k}{\Delta}  \left( \left[ \sigma_z^{(j)}, \left(\sigma_{x}^{(j)} \sigma_{x}^{(k)} + \sigma_{y}^{(j)} \sigma_{y}^{(k)} \right) \right] + \left[ \sigma_z^{(k)}, \left(\sigma_{x}^{(j)} \sigma_{x}^{(k)} + \sigma_{y}^{(j)} \sigma_{y}^{(k)} \right) \right]  \right) \nonumber \\
    &= \sum_{jk} \frac{g_j g_k}{\Delta}  \left( \left[ \sigma_z^{(j)}, \sigma_{x}^{(j)} \right] \sigma_{x}^{(k)} + \left[ \sigma_z^{(j)},  \sigma_{y}^{(j)} \right] \sigma_{y}^{(k)} +  \sigma_{x}^{(j)} \left[ \sigma_z^{(k)}, \sigma_{x}^{(k)} \right] + \sigma_{y}^{(j)} \left[ \sigma_z^{(k)}, \sigma_{y}^{(k)} \right]  \right) \nonumber \\
    &= \sum_{jk} 2 i \frac{g_j g_k}{\Delta}  \left( \sigma_{y}^{(j)} \sigma_{x}^{(k)} - \sigma_{x}^{(j)} \sigma_{y}^{(k)} + \sigma_{x}^{(j)} \sigma_{y}^{(k)} - \sigma_{y}^{(j)} \sigma_{x}^{(k)}  \right)  = 0 \nonumber 
\end{align}

 \section{Deriving the expression of quantum violation of the NDC in ideal case }\label{app:analytical_section}

The analytical derivation of the violation is presented in details for a even integer spin $j$. Subsequently, the odd integer and half-integer cases are described. The probabilities and the NDC violations are summarised in Table~\ref{table1}. 

The spin is initially in the \textit{ground} state $\Ket{\Psi (0)}_s = \Ket{j,  -j}_s$. With the on-resonance Hamiltonian of Eq.~(\ref{eq:hamiltonian_reson}), a rotation  of angle $\phi = 2 g \alpha t_R$ is performed on the spin. Mathematically, the rotation (unitary) operator considered is $\mathcal{R}\left(\phi \right) = e^{-i \phi J_y}$. After the rotation, the spin is described by the CS in Eq.~(\ref{eq:spin_cs}). Next, the cavity and the spins are let to interact for a time interval $t_I$, with the cavity out of resonance, with the Hamiltonian in Eq.~(\ref{eq:hamiltonian_off}). The input cavity quantum state is taken in be a CS:  $\Ket{\alpha_0}_c = e^{- \frac{1}{2} |\alpha_0|^2} \sum_n \frac{\alpha_0^n}{\sqrt{n!} } \Ket{n}_c $ with $\alpha_0 = \frac{1+i}{\sqrt{2}}|\alpha_0|$, where $|\alpha_0| \gg 1$. After the interaction, the entangled state between the cavity and the spin system is
\begin{equation}
\label{eq:spin_cavity_state}
    \Ket{\Psi(t_I +t_R)}_{sc} = \sum_{m=-j}^{j}d^{(j)}_{m, -j} \left( \phi \right) e^{ i 2 \chi m^2 t }  \ket{j,m}_s  \Ket{ \alpha_0 e^{- i \chi t_I m }  }_c  = \ket{\phi_e}_{s} \Ket{ \alpha_0}_c + \ket{\phi_o}_{s} \Ket{- \alpha_0}_c,
\end{equation} 
up to a global phase (proportional to $J^2$), and where the second equality follows from the choice $ \chi t_I = \pi$, where we note that at this time the phase is trivially zero.  

The projective measurement $\{\Pi_+, \Pi_-\}$ defined in Eq.~(\ref{projectionsmain}) are performed on the cavity field, resulting in the following unnormalized post-measurement (UPM) states,
\begin{align*}
    & \Ket{\Psi^+}^{\text{U}}_{sc} := \Pi_+ \Ket{\Psi(t_I +t_R)}_{sc} =  \ket{\phi_e}_{s} \frac{1}{\pi^{1/4}} \int_0^{\infty} dx  e^{ -\frac{1}{2} \left(x - |\alpha_0| \right)^2 } e^{i x |\alpha_0|} \ket{x}_c  +  \ket{\phi_o}_{s} \frac{1}{\pi^{1/4}} \int_0^{\infty} dx  e^{ -\frac{1}{2} \left(x + |\alpha_0| \right)^2} e^{-i x |\alpha_0|} \ket{x}_c,\\
    & \Ket{\Psi^-}^{\text{U}}_{sc} := \Pi_- \Ket{\Psi(t_I +t_R)}_{sc} =  \ket{\phi_e}_{s} \frac{1}{\pi^{1/4}} \int_{-\infty}^0 dx  e^{ -\frac{1}{2} \left(x - |\alpha_0| \right)^2 } e^{i x |\alpha_0|} \ket{x}_c  +  \ket{\phi_o}_{s} \frac{1}{\pi^{1/4}} \int_{-\infty}^0 dx  e^{ -\frac{1}{2} \left(x + |\alpha_0| \right)^2} e^{-i x |\alpha_0|} \ket{x}_c,
\end{align*}
where the position representation of the coherent state was used $\braket{x|\alpha}_c = \pi^{-1/4} e^{-{\frac{1}{2}\left(x-\sqrt{2} \text{Re}(\alpha) \right)^2} + i \sqrt{2} x \text{Im}(\alpha) }$. Here ``U''  denotes unnormalized states.

The two Gaussian distributions have mean at $x=\pm |\alpha_0|$ and standard deviation being equal to $1$.  
Since, as assumed earlier, $|\alpha_0| \gg 1$, the Gaussian distribution with mean at $x=+|\alpha_0|$ approximately vanishes in the region from $x \in (-\infty, 0)$ (and similarly  in the region: $x \in (0, \infty)$ for the case of the Gaussian distribution with mean at $x=-|\alpha_0|$).
Hence, we can approximate that 
\begin{equation*}
    \Ket{\Psi^+}^{\text{U}}_{sc}
    \approx \ket{\phi_e}_{s} \frac{1}{\pi^{1/4}}   \int_0^{\infty} dx  e^{ -\frac{1}{2}\left(x - |\alpha_0| \right)^2 } e^{i x |\alpha_0|} \ket{x}_c, \hspace{1.0cm} \Ket{\Psi^-}^{\text{U}}_{sc} \approx  \ket{\phi_o}_{s} \frac{1}{\pi^{1/4}}  \int_{-\infty}^{0} dx  e^{ -\frac{1}{2}\left(x + |\alpha_0| \right)^2 } e^{-i x |\alpha_0|}\ket{x}_c.
\end{equation*} 
With the similar approximation, and tracing out the cavity, the UPM spin states are $\Ket{\Psi^+}^{\text{U}}_s
\approx \ket{\phi_e}_{s}$ and 
$ \Ket{\Psi^-}^{\text{U}}_s \approx  \ket{\phi_o}_{s}$. 

For even integer $j$, it is possible to note that
\begin{equation}
    \Ket{\phi}_s + \Ket{-\phi}_s =  \sum_{m=-j}^{j} \left( d_{m, -j}^{(j)} \left( \phi \right) + \left( - 1 \right)^{m+j} d_{m, -j}^{(j)} \left(\phi \right) \right)\ket{j,m}_s = 2 \ket{\phi_e}_{s},
    \label{rewrite_even_odd}
\end{equation}
 where the following properties were used
$ d_{m, -j}^{(j)} \left(- \phi \right) = d_{-j, m}^{(j)} \left( \phi \right) = \left( - 1 \right)^{m+j}d_{m, -j}^{(j)} \left( \phi \right).$
Similarly,
$\Ket{\phi}_s - \Ket{-\phi}_s = 2 \ket{\phi_o}_{s}$. Hence, the UPM spin states can be written as
\begin{equation}
\label{eq:UPM_first_ideal}
\Ket{\Psi^{\pm}}^{\text{U}}_s = \frac{1}{2} \left( \Ket{\phi}_s \pm \Ket{-\phi}_s  \right).
\end{equation}
Using the facts that $\mathcal{R}^{\dag}\left(-\phi \right) =  \mathcal{R}\left(\phi \right)$ and  $\mathcal{R}\left(\phi_1 \right)\mathcal{R}\left(\phi_2 \right) =  \mathcal{R}\left(\phi_1 + \phi_2 \right)$ , the inner product $\Braket{-\phi | \phi}_s = d_{-j, -j}^{(j)} \left( 2 \phi \right) $. The probabilities of the outcomes are then given by the norm of the UPM states:
\begin{equation}
   P_1(\pm) = \braket{\Psi^{\pm}| \Psi^{\pm}}^{\text{U}}_s = \frac{1}{2} \pm \frac{1}{2} d_{-j, -j}^{(j)} \left( 2 \phi \right) =  \frac{1}{2} \pm \frac{1}{2}  (\cos\phi)^{2j} .
   \label{prob_pm_even}
\end{equation}


A similar analysis can be applied to the second rotation, which gives the rotated UPM states
\begin{equation*}
    \Ket{\Psi^{\pm} (t_R)}^{\text{U}}_s :=  \mathcal{R}\left(\phi\right) \Ket{\Psi^{\pm}}^{\text{U}}_s =  \frac{1}{2} \left( \Ket{2 \phi}_s  \pm \Ket{j,-j}_s  \right).
\end{equation*}
The coupling with the out of resonance cavity in the same CS $\Ket{\alpha_0}_c$ and with same interaction time $t_I$ (recalling that $j$ is even) similarly split the cavity in $\ket{\pm \alpha_0}_c$. The joint state is then
\begin{align}
    \Ket{\Psi^{\pm} (t_R + t_I)}^{\text{U}}_{sc}
    &=  \frac{1}{2} \left(\pm \Ket{j,-j}_s  + \Ket{2 \phi}_{e_s}  \right) \Ket{\alpha_0}_c  + \frac{1}{2} \Ket{2 \phi}_{o_s}  \Ket{-\alpha_0}_c. 
\end{align}
Performing  the same projective measurement, tracing out the cavity Hilbert space and assuming  $|\alpha_0| \gg 1$, we have the following UPM spin states,
\begin{equation*}
    \Ket{\Psi^{\pm,+}(t_R + t_I)}^{\text{U}}_{s}=  \pm \frac{1}{2} \left( \Ket{j,-j}_s \pm \frac{1}{2} \left( \Ket{2 \phi}_s + \Ket{-2 \phi}_s \right) \right), \hspace{0.5cm} 
    \Ket{\Psi^{\pm,-}_{\text{PM}} (t_R + t_I)}^{\text{U}}_{s}  =   \frac{1}{4} \left( \Ket{2 \phi}_s - \Ket{-2 \phi}_s \right),
\end{equation*}
where $\Ket{\Psi^{i,j}(t_R + t_I)}^{\text{U}}_{s}$ denotes the UPM states of the spins when the first and the second measurements give the outcomes $i$ and $j$ respectively. Hence, the joint probabilities can be computed as follows
\begin{equation*}
    P_{1,2}(\pm +) = \frac{3}{8} \pm  \frac{1}{2}   d_{-j, -j}^{(j)} \left(  2\phi \right)  + \frac{1}{8} d_{-j, -j}^{(j)} \left( 4 \phi \right),\hspace{1cm} P_{1,2}\left( \pm - \right) =  \frac{1}{8} - \frac{1}{8} d_{-j, -j}^{(j)} \left( 4 \phi \right).
\end{equation*}

In order to find $P_{2}( \pm)$, we use the probabilities in Eq.~(\ref{prob_pm_even}) with $\phi \to 2 \phi$, as in this case, intermediate measurement is not implemented and only the two rotations are performed one after the other. The violations of the NDC for even integer $j$ are then computed according to Eq.~(\ref{eq:ndc}) and recover the result claimed in Eq.~(\ref{eq:violation_ideal}).

In order to extend this analysis to every $j$, similar proofs were found for odd and half-integer spins. For odd $j$, the only difference is that $\Ket{\phi}_s + \Ket{-\phi}_s = 2 \ket{\phi_o}_{s}$ and $\Ket{\phi}_s - \Ket{-\phi}_s = 2 \ket{\phi_e}_{s}$, leading to  $\Ket{\Psi^{\pm}}^{\text{U}}_{s} = \frac{1}{2} \left( \Ket{\phi}_s \mp \Ket{-\phi}_s  \right)$. The results can easily be  derived following the same proof for even $j$.

For spin half-integer, it is possible to rewrite $j=n+1/2$. This changes the entanglement between the cavity out of resonance and the spin in Eq.~(\ref{eq:spin_cavity_state}), according to 
\begin{equation*}
    \Ket{\Psi(t_R + t_I)}_{sc} =  \sum_{m=-n-1}^{n}d^{(j)}_{m+\frac{1}{2},-j} \left( \phi \right)  \ket{j, m+1/2}_s  \Ket{ \alpha_0 e^{- i \chi t_I  \left(m + \frac{1}{2}\right) }  }_c = \ket{\phi_e}_{s} \Ket{ - i \alpha_0}_c + \ket{\phi_o}_{s} \Ket{i \alpha_0}_c.
\end{equation*}
Here $\ket{j, m+1/2}$ is the eigenstate of $J_z$ with eigenvalue $m + 1/2$ for a spin-$j$ system with $j$ being half-integer, where $m$ is an integer and $m \in \{-n-1, -n, -n+1, \cdots, n\}$ with $n=j-1/2$. In this case, even and odd CS are defined as 
\begin{align}
    &\ket{\phi_e}_{s} := \sum_{m \in \text{even}} d_{m+\frac{1}{2}, -j}^{(j)} \left( \phi \right) \ket{j,m+1/2}_s, \hspace{0.5cm}\ket{\phi_o}_{s} := \sum_{m \in \text{odd}} d_{m+\frac{1}{2}, -j}^{(j)} \left( \phi \right) \ket{j,m+1/2}_s.\nonumber 
\end{align} 
Note, the $i$ difference in the cavity states. This motivated the choice of $\alpha_0 = \frac{1+i}{\sqrt{2}}|\alpha_0|$, ensuring that also for half-integer $j$, the projective measurement gives the UPM states $\Ket{\Psi^+}^{\text{U}}_{s}
\approx \ket{\phi_e}_{s}$ and 
$ \Ket{\Psi^-}^{\text{U}}_{s} \approx  \ket{\phi_o}_{s}$.  The probabilities follows from the norm. 

\newcolumntype{M}[1]{>{\centering\arraybackslash}m{#1}}
\newcolumntype{N}{@{}m{0pt}@{}}

\begin{table}[!t]
  \begin{center}
\begin{tabular}{ |M{1.8cm}||M{3.5cm}|M{3.5cm}|M{3.5cm}|M{3.5cm}| N } \hline
   \vspace{10pt}  &  Integer $j$  & Half-Integer $j = n + \frac{1}{2}$  &  Integer $j$  & Half-Integer $j = n + \frac{1}{2}$  \\ [10pt] \hline
   \vspace{5pt}  &   even $j$ & odd $n$  &  odd $j$ & even $n$ \\ [5pt] \hline \hline 
 \vspace{5pt} $ P_2(\pm) $   &  
 \multicolumn{2}{c|}{$ \frac{1}{2} \pm \frac{1}{2}  \left( \cos 2 \phi \right)^{2j}  $}  & 
 \multicolumn{2}{c|}{$ \frac{1}{2} \mp \frac{1}{2}  \left( \cos 2 \phi \right)^{2j} $}  & \\  [5pt] \hline

 \vspace{5pt} $ P_{1,2}(\pm +) $   &  
 \multicolumn{2}{c|}{$ \frac{3}{8} \pm  \frac{1}{2}   \left( \cos \phi \right)^{2j}  + \frac{1}{8} \left( \cos 2 \phi \right)^{2j}  $ } & 
 \multicolumn{2}{c|}{$ \frac{1}{8} - \frac{1}{8} \left( \cos 2 \phi \right)^{2j}  $ }  & \\ [5pt]  \hline

  \vspace{5pt}$ P_{1,2}(\pm -) $   &  
 \multicolumn{2}{c|}{$ \frac{1}{8} - \frac{1}{8} \left( \cos 2 \phi \right)^{2j}  $ } & 
 \multicolumn{2}{c|}{$ \frac{3}{8} \mp  \frac{1}{2}   \left( \cos \phi \right)^{2j}   + \frac{1}{8} \left( \cos 2 \phi \right)^{2j}  $} & \\ [10pt] \hline 
 
\vspace{5pt} $ V_{\pm} $   &  
 \multicolumn{2}{c|}{$ \mp \frac{1}{4} \left[1- (\cos \left( 2 \phi \right))^{2j}   \right]$}  & 
 \multicolumn{2}{c|}{$ \pm \frac{1}{4} \left[1- (\cos \left( 2 \phi \right))^{2j}   \right]  $} & \\  [5pt] 
 \hline 
\end{tabular}
\caption{Expressions for probability distributions and the violations of NDC for different $j$.} \label{table1}
\end{center}
\end{table} 

\section{Deriving the expression of the violation of the NDC when the two arbitrary rotation angles $\phi_1$ and $\phi_2$}\label{app2}

Let us consider $j$ to be integer. After the first rotation $\mathcal{R}\left(\phi_1 \right)$ and first measurement 
(i.e., at time $t_R^{(1)} + t_I$), the UPM spin states are 
\begin{equation}
\Ket{\Psi^{+}}^{\text{U}}_s = \ket{\phi_1}_{e_s}, \hspace{0.5cm} \Ket{\Psi^{-}}^{\text{U}}_s = \ket{\phi_1}_{o_s}.
\label{UPMapp2}
\end{equation}
After the second rotation $\mathcal{R}\left(\phi_2 \right)$, the UPM spin states  evolve to:
\begin{align*}
  &  \Ket{\Psi^+\left(t_R^{(2)}\right)}^{\text{U}}_s 
  =\sum_{m \in \text{even}}   d^{(j)}_{m, -j} \left( \phi_1 \right)  \sum_{m' = - j}^j  d^{(j)}_{m',m}\left( \phi_2 \right) \ket{j,m'}_s, \\
  & \Ket{\Psi^-\left(t_R^{(2)}\right)}^{\text{U}}_s 
  =\sum_{m \in \text{odd}}   d^{(j)}_{m, -j} \left( \phi_1 \right)  \sum_{m' = - j}^j  d^{(j)}_{m',m}\left( \phi_2 \right) \ket{j,m'}_s.
\end{align*}

Next, the spin system and the cavity in the aforementioned CS $\Ket{\alpha_0}$ undergoes the same off-resonance interaction as described earlier. After the interaction, the spin-cavity entangled state (when $+$ outcome is obtained in the first measurement) is 
\begin{equation*}
    \Ket{\Psi^+ \left(t_R^{(2)} + t_I\right)}^{\text{U}}_{sc} = \sum_{m \in \text{even}}   d^{(j)}_{m, -j} \left( \phi_1 \right)  \sum_{m' = - j}^j   d^{(j)}_{m',m}\left( \phi_2 \right) \ket{j,m'}_s \Ket{ \left( - 1 \right)^{m'}\alpha_0 }_c.
\end{equation*}
Subsequently, the aforementioned projective measurement is performed on the cavity field. When $+$ outcome is obtained in this second measurement, the UPM joint state of the spin-cavity becomes
\begin{align*}
    \Ket{\Psi^{++}(t_R^{(2)} + t_I)}^{\text{U}}_{sc} \approx \frac{1}{\pi^{1/4}} \sum_{m, m' \in \text{even}}   d^{(j)}_{m, -j}(\phi_1)    d^{(j)}_{m', m}(\phi_2) \ket{j,m'}_s  \int_0^{\infty} dx  e^{ -\frac{1}{2} \left(x - |\alpha_0| \right)^2 } e^{i x |\alpha_0|}\ket{x}_c.
\end{align*} 
where the same approximation ensured by $|\alpha_0| \gg 0$ was taken.
The norm of the above UPM gives the following joint probability:
\begin{equation*}
        P_{1,2}(++) \approx \sum_{m, m' \in \text{even}} d^{(j)}_{m, -j}(\phi_1) d^{(j)}_{m', -j}(\phi_1) \left( \sum_{k \in \text{even}}   d^{(j)}_{k,m}(\phi_2)    d^{(j)}_{k,m'}(\phi_2) \right).
\end{equation*}
The other joint probabilities can be calculated similarly. The probabilities $P_2(\pm)$ can be obtained from the norms of the states (\ref{UPMapp2}) with $\phi_1$ being replaced by $\phi_1 + \phi_2$.

{\centering
	\begin{table}[t]
		\begin{tabular}{ |c|c| } 
			\hline
			   & Integer $j$  \\
      \hline
			  $P_2(+)$ & $\sum\limits_{m \in \text{even}}  \left| d^{(j)}_{m, -j}\left( \phi_1 + \phi_2 \right) \right|^2$ \\ \hline
			$P_2(-)$ & $\sum\limits_{m \in \text{odd}}  \left| d^{(j)}_{m, -j} \left( \phi_1 + \phi_2 \right) \right|^2$  \\\hline
			$P_{1,2}(+,+)$ & $\sum\limits_{m, m'' \in \text{even}}   d^{(j)}_{m, -j} (\phi_1) d^{(j)}_{m'', -j}(\phi_1)  \left(\sum\limits_{m'\in \text{even}} d^{(j)}_{m', m}(\phi_2) d^{(j)}_{m', m''}(\phi_2) \right)$ \\\hline
			$P_{1,2}(+,-)$ & $\sum\limits_{m, m'' \in \text{even}}   d^{(j)}_{m, -j}(\phi_1) d^{(j)}_{m'', -j}(\phi_1)  \left(\sum\limits_{m'\in \text{odd}} d^{(j)}_{m', m}(\phi_2)  d^{(j)}_{m', m''}(\phi_2)\right)$ \\\hline
			$P_{1,2}(-,+)$ & $\sum\limits_{m, m'' \in \text{odd}}   d^{(j)}_{m, -j}(\phi_1) d^{(j)}_{m'', -j}(\phi_1)  \left(\sum\limits_{m'\in \text{even}} d^{(j)}_{m', m}(\phi_2)  d^{(j)}_{m', m''}(\phi_2)\right)$ \\\hline
			$P_{1,2}(-,-)$ & $\sum\limits_{m, m'' \in \text{odd}}   d^{(j)}_{m, -j}(\phi_1) d^{(j)}_{m'', -j}(\phi_1)  \left(\sum\limits_{m'\in \text{odd}} d^{(j)}_{m', m}(\phi_2)  d^{(j)}_{m', m''}(\phi_2)\right)$ \\\hline
			
			& Half-integer $j$  \\
   \hline 
			  $P_2(+)$ & $\sum\limits_{m \in \text{even}}  \left| d^{(j)}_{m + \frac{1}{2}, -j}\left( \phi_1 + \phi_2 \right) \right|^2$ \\ \hline
			$P_2(-)$ & $\sum\limits_{m \in \text{odd}}  \left| d^{(j)}_{m + \frac{1}{2}, -j} \left( \phi_1 + \phi_2 \right) \right|^2$  \\ \hline
			$P_{1,2}(+,+)$ & $\sum\limits_{m, m'' \in \text{even}}   d^{(j)}_{m+\frac{1}{2}, -j}(\phi_1) d^{(j)}_{m''+\frac{1}{2}, -j}(\phi_1) \left( \sum\limits_{m' \in \text{even}}   d^{(j)}_{m'+\frac{1}{2}, m+\frac{1}{2}}(\phi_2)     d^{(j)}_{m'+\frac{1}{2}, m''+\frac{1}{2}}(\phi_2)  \right)$ \\ \hline
			$P_{1,2}(+,-)$ & $\sum\limits_{m, m'' \in \text{even}}   d^{(j)}_{m+\frac{1}{2}, -j}(\phi_1) d^{(j)}_{m''+\frac{1}{2}, -j}(\phi_1) \left( \sum\limits_{m' \in \text{odd}}   d^{(j)}_{m'+\frac{1}{2}, m+\frac{1}{2}}(\phi_2)     d^{(j)}_{m'+\frac{1}{2}, m''+\frac{1}{2}}(\phi_2)  \right)$ \\ \hline
			$P_{1,2}(-,+)$ & $\sum\limits_{m, m'' \in \text{odd}}   d^{(j)}_{m+\frac{1}{2}, -j}(\phi_1) d^{(j)}_{m''+\frac{1}{2}, -j}(\phi_1) \left( \sum\limits_{m' \in \text{even}}   d^{(j)}_{m'+\frac{1}{2}, m+\frac{1}{2}}(\phi_2)     d^{(j)}_{m'+\frac{1}{2}, m''+\frac{1}{2}}(\phi_2)  \right)$ \\ \hline
			$P_{1,2}(-,-)$ & $\sum\limits_{m, m'' \in \text{odd}}  d^{(j)}_{m+\frac{1}{2}, -j}(\phi_1) d^{(j)}_{m''+\frac{1}{2}, -j}(\phi_1) \left( \sum\limits_{m' \in \text{odd}}   d^{(j)}_{m'+\frac{1}{2}, m+\frac{1}{2}}(\phi_2)     d^{(j)}_{m'+\frac{1}{2}, m''+\frac{1}{2}}(\phi_2)  \right)$ \\ \hline
			\hline
		\end{tabular}
		\caption{Expressions for the probabilities with different $j$. For integer $j$, $m, m''\in \{-j, -j+1, \cdots, j-1, j\}$. On the other hand, for half-integer $j$, $m, m''\in \{-j-1/2, -j-1/2+1, \cdots, j-1/2\}$.  
  } \label{tab2}
	\end{table}
} 

The above calculation can also be extended for half-integer $j$ easily. The expressions for the probabilities (with integer as well as half-integer $j$) are summarized in Table~\ref{tab2}. From these expressions, violations of the NDC is calculated as function of $\phi_1$ and $\phi_2$.

\section{Analytical Solution for $\phi_2 = \frac{\pi}{2}-  \phi_1 $}  \label{app:error_rotation}

In this appendix, the first rotation is of angle $\phi_1 = \phi$ and the second $\phi_2 = \frac{\pi}{2}-  \phi_1 $. As an example, we will consider $j$ to be even and compute $V_+$. The results are easily extendable to other $j$ and to $V_-$. After the first measurement the UPM state will be described by Eq.~(\ref{eq:UPM_first_ideal}). Hence, applying the second rotation: 
\begin{equation*}
    \Ket{\Psi^{\pm} (t_R)}^{\text{U}}_s := \mathcal{R}\left( \frac{\pi}{2} - \phi \right)  \Ket{\Psi^{\pm}}^{\text{U}}_s =  \frac{1}{2} \left( \Ket{\frac{\pi}{2}}_s \pm \Ket{\frac{\pi}{2} - 2 \phi}_s   \right).
\end{equation*}
After the second measurement, the UPM state will be:
\begin{equation*}
    \Ket{\Psi^{\pm +} }^{\text{U}}_s =  \frac{1}{2} \left( \Ket{\frac{\pi}{2}}_{e_s} \pm \Ket{\frac{\pi}{2} - 2 \phi}_{e_s}   \right) = \frac{1}{4} \left( \Ket{\frac{\pi}{2}}_s +  \Ket{- \frac{\pi}{2}}_s \pm \Ket{\frac{\pi}{2} - 2 \phi}_s  \pm \Ket{- \frac{\pi}{2} + 2 \phi}_s  \right), 
\end{equation*}
which has norm:
\begin{align*}
     P_{1,2}(\pm +) &= \frac{1}{8} \Big( 2 + \Re \left[ \Braket{\frac{\pi}{2} \big| - \frac{\pi}{2} }_s \right] \pm  \Re \left[ \Braket{\frac{\pi}{2} \big| \frac{\pi}{2} - 2 \phi }_s \right] \pm  \Re \left[ \Braket{\frac{\pi}{2} \big|-  \frac{\pi}{2} + 2 \phi }_s \right]  \\ 
     &  \hspace{1cm} \pm  \Re \left[ \Braket{- \frac{\pi}{2} \big| \frac{\pi}{2} - 2 \phi }_s \right] \pm  \Re \left[ \Braket{- \frac{\pi}{2} \big|- \frac{\pi}{2} + 2 \phi }_s \right] + \Re \left[ \Braket{\frac{\pi}{2} - 2 \phi  \big|- \frac{\pi}{2} + 2 \phi }_s \right]  \Big),
\end{align*}
where $\Re$ implies the real part, and it will act trivially. Note that a general inner product can be simplified as $\braket{\phi_1|\phi_2}_s = \braket{-j |\mathcal{R}^\dag \left( \phi_1 \right) \mathcal{R} \left( \phi_2 \right)| -j}_s = \braket{-j |\mathcal{R} \left( \phi_2 - \phi_1 \right)| -j}_s = d_{-j,-j}^{(j)} \left( \phi_2 - \phi_1 \right)$. It implies that:
 \begin{equation*}
     P_{1,2}(\pm +) = \frac{1}{8} \left[ 2 + d_{-j,-j}^{(j)} \left(- \pi \right) \pm  d_{-j,-j}^{(j)} \left( - 2 \phi \right) \pm  d_{-j,-j}^{(j)} \left( - \pi + 2 \phi \right)   \pm  d_{-j,-j}^{(j)} \left( \pi - 2\phi \right) \pm  d_{-j,-j}^{(j)} \left( 2 \phi \right) + d_{-j,-j}^{(j)} \left( -\pi + 4 \phi \right) \right].
\end{equation*}
Hence, 
\begin{equation*}
    P_{1,2}(+ +) + P_{1,2}(-+)= \frac{1}{4} \left[ 2 + d_{-j,-j}^{(j)} \left(- \pi \right) + d_{-j,-j}^{(j)} \left( -\pi + 4 \phi \right) \right].
\end{equation*}
Recalling that 
\begin{equation*}
    d_{m',m}^{(j)} \left(- \phi \right) = \left( -1 \right)^{m'-m} d_{m',m}^{(j)} \left( \phi \right), \hspace{0.5cm}
    d_{m',m}^{(j)} \left( \pi \right) = \left( -1 \right)^{j-m} \delta_{m',-m},
    \hspace{0.5cm}
    d_{m',m}^{(j)} \left( \pi - \phi \right) = \left( -1 \right)^{j+m'} d_{m', -m}^{(j)} \left( \phi \right),
\end{equation*}
then it follows that  $P_{1,2}(+ +) + P_{1,2}(-+)= 1/2 + d_{-j,j}^{(j)} \left(4 \phi \right)/4$. When the first measurement is not performed, the total rotation angle is $\pi/2$, which implies that the UPM for the $+$ outcome is $\ket{\pi/2}_{e_s}$ and the probability
 \begin{equation*}
     P_{2}(+) = \frac{1}{4} \left( \Bra{\frac{\pi}{2 }}_s + \Bra{-\frac{\pi}{2 }}_s \right)\left( \Ket{\frac{\pi}{2 }}_s + \Ket{- \frac{\pi}{2 }}_s \right) = \frac{1}{2}\left( 1 + d_{-j,-j}^{(j)} \left(\pi \right) \right).
\end{equation*} 
Using the fact that $ d_{-j,-j}^{(j)} \left(\pi \right) = \left(\cos(\pi/2) \right)^{2j} = 0$ and $d_{-j,j}^{(j)} \left( \phi \right) = \sin^{2j}(\phi/2)$, the violation of Eq.~(\ref{eq:violation_error_rot_an}) follows.  Examples of the violations for different spin values is given in Fig.~\ref{fig:phi121}.

\section{Decoherence} \label{app:decoherence}

In order to compute the violation of the NDC with decoherence, first of all, the solution of the Master equation given by Eq.~(\ref{eq:master}) is found, and then the violation of MR under decoherence is computed.  The first term in Eq.~(\ref{eq:master}) gives the unitary evolution previously used to entangle the cavity and the spin: $\rho(t)_{sc} = U(t) \rho(0)_{sc} U^\dag(t)$, where $U(t) = \exp{\left(- i t \chi a^\dag a  J_z \right)}$. Under the unitary evolution, the density matrix at time $t$ is generally given by
\begin{equation}
    \rho(t)_{sc} = \sum_{m,n} \rho_{m,n} \ket{m}\bra{n}_{s} \otimes \ket{\alpha_m(t)}   \bra{\alpha_n(t)}_c \;,
    \label{dmunderdecoherence}
\end{equation}
where $\rho_{m,n}$ are the elements of the density matrix, and $\ket{\alpha_m(t)}_c = \ket{e^{- i m \chi t} \alpha_m}_c$. Hence, it is possible to break the problem into solving the full dynamics  for each $\rho_{m,n}\ket{\alpha_m(t)}   \bra{\alpha_n(t)}_c$ element. The additional terms in Eq.~(\ref{eq:master}) were studied in Ref.~\cite{dissipation}, where it was concluded that under cavity leaking, the coherent states basis evolves according to 
\begin{equation}
    \ket{\alpha} \bra{\beta}_c \to \braket{\alpha|\beta}^{1-e^{-\gamma_c t}}\ket{\alpha e^{-\gamma_c t/2}} \bra{\beta e^{-\gamma_c t/2}}_c,
    \label{dissipation}
\end{equation} 
where $\ket{\alpha}_c,\; \ket{\beta}_C$ are two cavity CSs.

Now, to compute the full dynamics, we should consider that each $\alpha_m$ is a function of $\gamma_c$ in addition to $t$. Let us consider infinitesimal time step $dt$. On one hand, at every $dt$, the unitary evolution and the dissipation will update the amplitude of the $m$ CS according to
\begin{equation*}
    \alpha_m(\gamma_c, t) \to \alpha_m(\gamma_c, t) \left(1 - i m \chi dt \right), \hspace{1.5cm}  \alpha_m(\gamma_c, t) \to \alpha_m(\gamma_c, t) \left(1 - \frac{\gamma_c}{2} dt \right), 
\end{equation*}
respectively. Combining the two, the following differential equation can be formulated and solved
\begin{equation}
    \frac{\partial \alpha_m(\gamma_c, t) }{\partial t} = - \left(i m \chi + \frac{\gamma_c}{2} \right) \alpha_m(\gamma_c, t) \;\;\; \implies \;\;\; \alpha_m(\gamma_c, t) = \alpha_0 e^{- \left(i m \chi + \gamma_c/2 \right) t} ,
    \label{eq:alpha_decohered}
\end{equation}
where the boundary condition $\alpha_m(\gamma_c, 0) = \alpha_0 $ was imposed $\forall m$. On the other hand, it is possible to note that the coefficients $\rho_{m,n}(\gamma_c, t)$ will not be effected by the unitary evolution, and only by decoherence following Eq.~(\ref{dissipation}).  At every $dt$,
\begin{equation*}
    \rho_{m,n}(\gamma_c, t) \to \braket{\alpha_m(\gamma_c, t) | \alpha_n(\gamma_c, t) }^{\gamma_c dt} \rho_{m,n}(\gamma_c, t). 
\end{equation*}

Hence, evolving $\rho_{m,n}(\gamma_c, 0)$, 
\begin{equation*}
    \rho_{m,n}(\gamma_c, t)  = \rho_{m,n}(\gamma_c, 0)\exp{\left(- D_{m,n}(\gamma_c, t) \right)},
\end{equation*}
where, using $\braket{\alpha | \beta} = \exp{ \frac{-1}{2}(|\alpha|^2 + |\beta|^2 - 2 \alpha^* \beta)}$, we get
\begin{equation*}
    D_{m,n}(\gamma_c, t)  =  \frac{\gamma_c}{2} \int_0^t dt' \left(|\alpha_m(\gamma_c, t')|^2 + |\alpha_n(\gamma_c, t')|^2 - 2 \alpha_m(\gamma_c, t')  \alpha_n^*(\gamma_c, t') \right).
\end{equation*} 
Using (\ref{eq:alpha_decohered}) and evaluating the integral, Eq.~(\ref{eq:decoherence_leaking}) is recovered.

Spin dephasing, of decoherence rate $\gamma_s$, can be included in the model by replacing
\begin{equation}
    \ket{j,m} \bra{j,n}_s \to e^{-\gamma_s t (m-n)^2} \ket{j,m} \bra{j,n}_s.
\end{equation}
Hence,

In order to keep the discussion general and do not commit to a physical system, we will compute the effect of the decoherence in the measurement probabilities in terms of the ratios $r_c = \gamma_c/\chi$ and $r_s = \gamma_s/\chi$.

At first, we assume that $j$ is integer. Substituting $t= t_I = \pi/\chi$ in Eq.~(\ref{eq:decoherence_dens}), the density matrix is
\begin{equation*}
    \rho(t_I)_{sc} = \sum_{m,n =-j}^j d^{(j)}_{m, -j} \left(\phi \right) d^{(j)}_{n, -j}  \left(\phi \right) e^{-(m-n)^2 \pi r_s } e^{-D_{m,n}(r_c) } \ket{j,m} \bra{j,n}_s \otimes \Ket{ (-1)^{m} \alpha_0 e^{ - r_c \pi/2 } } \Bra{ (-1)^{n} \alpha_0 e^{ - r_c \pi / 2   } }_c, 
\end{equation*}
where
\begin{equation*}
    D_{m,n}( r_c)  := D_{m,n}( \gamma_c, t_I) = |\alpha_0|^2 \left( 1 - e^{-\pi r_c} + \frac{(-1)^{m-n} e^{-\pi r_c} - 1}{1+\frac{i}{r_c} (m-n)}\right).
\end{equation*}
Hence, taking the projection measurement and tracing out the cavity, the UPM spin state is given by:
\begin{equation*}
    \rho_{{\pm}_s}^{\text{U}}(r_s, r_c) = \sum_{m,n =-j}^j d^{(j)}_{m, -j}  \left(\phi \right) d^{(j)}_{n, -j}   \left(\phi \right) e^{-(m-n)^2  \pi r_s} e^{-D_{m,n}( r_c) } \mathcal{P}_{m,n}^{(\pm)} (r_c) \ket{j,m} \bra{j,n}_s,
\end{equation*}
where
\begin{equation*}
    \mathcal{P}_{m,n}^{(\pm)}(r_c) = \Braket{ (-1)^{n}  \alpha_0 e^{ - r_c \pi /2 } |\Pi_\pm| (-1)^{m}\alpha_0 e^{- r_c \pi /2  }  }. 
\end{equation*}
The second rotation of the spin state will not be effected by previous decoherence and the same analysis can be applied to the second measurement. Hence, the UPM spin state,  when the outcome $\pm$ is obtained in the first measurement and the outcome $+$ is obtained in the second measurement, is given by,
\begin{align*}
    \rho_{{\pm +}_s}^{\text{U}}(r_s, r_c) &\approx \sum_{m,n=-j}^j d^{(j)}_{m, -j} \left(\phi \right) d^{(j)}_{n, -j}   \left(\phi \right) e^{- (m-n)^2 \pi r_s} e^{-D_{m,n}( r_c) } \mathcal{P}_{m,n}^{(\pm)}(r_c)  \\
    & \hspace{1cm} \cdot \sum_{m',n'=-j}^j d^{(j)}_{m', m}  \left(\phi \right) d^{(j)}_{n', n}  \left(\phi \right)  e^{- (m'-n')^2\pi r_s } e^{-D_{m',n'}( r_c) } \mathcal{P}_{m',n'}^{(+)}(r_c) \ket{j,m'} \bra{j,n'}_s.
\end{align*}
Taking the trace, the sum of the joint probabilities becomes
\begin{align*}
    P_{1,2}(+ +) + P_{1,2}(- +) &\approx  \sum_{m,n=-j}^j d^{(j)}_{m, -j} \left(\phi \right) d^{(j)}_{n, -j} \left(\phi \right)  e^{- (m-n)^2\pi r_s } e^{-D_{m,n}( r_c) } \left( \mathcal{P}_{m,n}^{(+)}(r_c) + \mathcal{P}_{m,n}^{(-)}(r_c) \right) \\
    & \hspace{2cm} \cdot \sum_{k=-j}^j d^{(j)}_{k, m} \left(\phi \right) d^{(j)}_{k, n}  \left(\phi \right) \mathcal{P}_{k,k}^{(+)}(r_c)  ,
\end{align*}
where $D_{k,k}(r_c) = 0 \forall k$ is used. Furthermore, we note that
\begin{align*}
     \mathcal{P}_{m,n}^{(+)}(r_c) + \mathcal{P}_{m,n}^{(-)}(r_c) &= \frac{1}{\sqrt{\pi}}
    \int_{-\infty}^{\infty} dx  \exp{ \left[ -\frac{1}{2} \left( \left(x - (-1)^{m} \alpha_0(r_c)   \right)^2 +  \left(x - (-1)^{n}\alpha_0(r_c)  \right)^2 \right) + i x \alpha_0(r_c)  \left((-1)^{m} - (-1)^{n}\right) \right]} \\
    &= \exp \left(\left(\frac{1}{2}+\frac{i}{2}\right) \alpha_0(r_c)  ^2 \left((-1)^n+i (-1)^m\right) \left((-1)^m+(-1)^{n+1}\right)\right),
\end{align*}
and 
\begin{equation*}
     \mathcal{P}_{k,k}^{(+)}(r_c) = \frac{1}{\sqrt{\pi}}
    \int_{0}^{\infty} dx  \exp{ \left[ -  \left(x - (-1)^{k} \alpha_0(r_c)   \right)^2  \right]} = \frac{1}{2}\left(1 + \text{Erf}\left( (-1)^{k} \alpha_0(r_c) \right) \right),
\end{equation*}
where $\alpha_0(r_c) = |\alpha_0| e^{- \frac{r_c \pi }{2} }$ is defined for conciseness.

Finally, when the first measurement is not performed, only the spin decoherence is present, i.e., during the time duration that the first measurement would have meant to happen, the spin state will decohere. This gives the spin state before the second rotation
\begin{equation}
    \rho_s(r_s) = \sum_{m,n =-j}^j d^{(j)}_{m, -j}\left(\phi \right) d^{(j)}_{n, -j}\left(\phi \right)  e^{-(m-n)^2 \pi r_s } \ket{j,m} \bra{j,n}_s .
\end{equation}
Applying the second rotation and the second measurement, the UPM spin-state (when the outcome $+$ is obtained) is given by,
\begin{align*}
    \rho_{+_s}^{\text{U}}(r_s, r_c) &\approx \sum_{m,n =-j}^{j} d^{(j)}_{m, -j} \left(\phi \right) d^{(j)}_{n, -j}   \left(\phi \right) e^{- (m-n)^2\pi r_s}  \\
    & \hspace{1cm} \cdot \sum_{m',n'=-j}^j d^{(j)}_{m', m}  \left(\phi \right) d^{(j)}_{n', n}  \left(\phi \right)  e^{- (m'-n')^2\pi r_s } e^{-D_{m',n'}( r_c) } \mathcal{P}_{m',n'}^{(+)}(r_c) \ket{j,m'} \bra{j,n'}_s.
\end{align*}

Taking the trace, we get the following,
\begin{equation}
    P_{2}(+) \approx   \sum_{m,n =-j}^j d^{(j)}_{m, -j}\left(\phi \right) d^{(j)}_{n, -j}\left(\phi \right)  e^{-(m-n)^2 \pi r_s } \sum_{k =-j}^j d^{(j)}_{k, m}\left(\phi \right) d^{(j)}_{k, n}  \left(\phi \right) \mathcal{P}_{k,k}^{(+)}(r_c).
\end{equation}

On the other hand, for half-integer $j$, following a similar calculation, it can be shown that
\begin{align*}
    P_{1,2}(+ +) + P_{1,2}(- +) &\approx  \sum_{m,n=-j-\frac{1}{2}}^{j-\frac{1}{2}} d^{(j)}_{m+\frac{1}{2}, -j} \left(\phi \right) d^{(j)}_{n+\frac{1}{2}, -j} \left(\phi \right)  e^{- (m-n)^2\pi r_s } e^{-D_{m,n}( r_c) } \left( \mathcal{Q}_{m,n}^{(+)}(r_c) + \mathcal{Q}_{m,n}^{(-)}(r_c) \right) \\
    & \hspace{2cm} \cdot \sum_{k=-j-\frac{1}{2}}^{j-\frac{1}{2}} d^{(j)}_{k+\frac{1}{2}, m+\frac{1}{2}} \left(\phi \right) d^{(j)}_{k+\frac{1}{2}, n+\frac{1}{2}}  \left(\phi \right) \mathcal{P}_{k,k}^{(+)}(r_c)  ,
\end{align*}
with
\begin{align*}
     \mathcal{Q}_{m,n}^{(+)}(r_c) + \mathcal{Q}_{m,n}^{(-)}(r_c) &= \frac{1}{\sqrt{\pi}}
    \int_{-\infty}^{\infty} dx  \exp{ \left[ -\frac{1}{2} \left( \left(x - (-1)^{m} \alpha_0(r_c)   \right)^2 +  \left(x - (-1)^{n}\alpha_0(r_c)  \right)^2 \right) + i x \alpha_0(r_c)  \left((-1)^{n} - (-1)^{m}\right) \right]} \\
    &= \exp \left(\left(\frac{1}{2}+\frac{i}{2}\right) \alpha_0(r_c)  ^2 \left((-1)^m+i (-1)^n\right) \left((-1)^n+(-1)^{m+1}\right)\right),
\end{align*}
and 
\begin{equation}
    P_{2}(+) \approx   \sum_{m,n =-j-\frac{1}{2}}^{j-\frac{1}{2}} d^{(j)}_{m+\frac{1}{2}, -j}\left(\phi \right) d^{(j)}_{n+\frac{1}{2}, -j}\left(\phi \right)  e^{-(m-n)^2 \pi r_s } \sum_{k =-j-\frac{1}{2}}^{j-\frac{1}{2}} d^{(j)}_{k+\frac{1}{2}, m+\frac{1}{2}}\left(\phi \right) d^{(j)}_{k+\frac{1}{2}, n+\frac{1}{2}}  \left(\phi \right) \mathcal{P}_{k,k}^{(+)}(r_c).
\end{equation}

\begin{figure}
    \centering
\includegraphics[width=0.9 \textwidth]{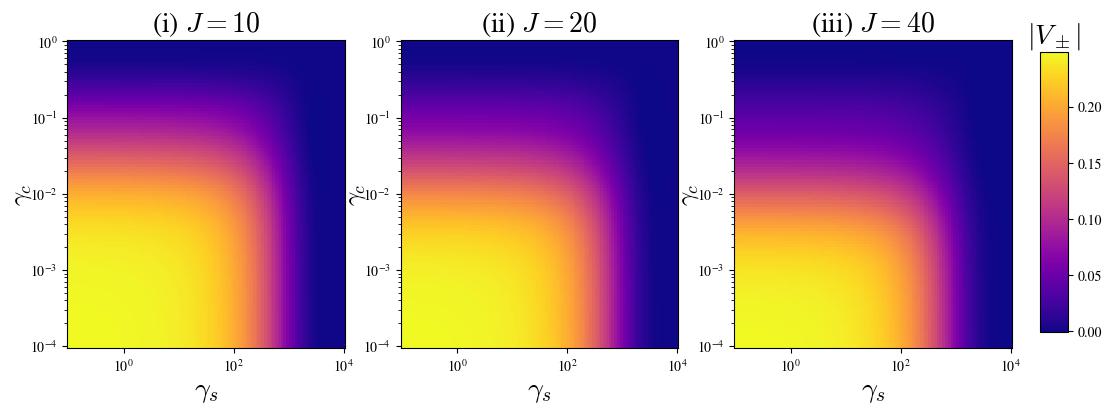}
    \caption{Decoherence effects on the violation of NDC for different values of the total spin $j$ as function of both the cavity leaking rate $\gamma_c$ and the spin decoherence rate $\gamma_s$, for  $\phi =\pi/4$.}
    \label{fig:dec_2d}
\end{figure}

The violation for different decoherence rates were numerically computed and the plot is given in Figs.~\ref{fig:dec_spin_caity} and~\ref{fig:dec_2d}, where we have taken $\phi =\pi/4$.

\section{Inhomogeneous Coupling} \label{app:inohomgeneus}

The initial \textit{ground} spin state can be rewritten as $\ket{1/2, -1/2}_s^{\otimes N} = \ket{j_s,-j_s}_s^{\otimes n}$. To keep the expression reasonably short, we will restrict the discussion to integer $j$ and $\alpha_0 \in \mathbb{R}$. The rotation and entanglement interaction times ($t_R$ and $t_I$) are chosen such that  
\begin{equation}
    \braket{\phi} := 2 \alpha_r \braket{g} t_R = \frac{\pi}{4},  \hspace{1.5cm}
    \braket{\theta} := \chi t_I = \pi.
\end{equation}
Furthermore, let us define the quantities $\phi_i := 2 \alpha_r \braket{g}_i t_R $ and $\theta_i := \braket{x}_i t_I  $.

The rotation Hamiltonian in Eq.~(\ref{inho_rot_ham}) will evolve the initial spin state to a product of CSs each with angle $\phi_i$ with respect to the quantization axis,
\begin{equation}
    \ket{\Psi (t_R)}_s = \bigotimes_{i=1}^{n} \Ket{\phi_i}_s =  \bigotimes_{i=1}^{n} \sum_{m_i = -j_s}^{j_s} d_{m_i, -j_s}^{(j_s)} \left( \phi_i \right) \Ket{j_s, m_i}_s := \sum_{\mathbf{m}} \mathcal{D}_{\mathbf{m}, -j_s}^{(j_s)} \Ket{j_s, \mathbf{m}}_s,
\end{equation}
where, for conciseness, the following notation was introduced: $\mathbf{m} = (m_1, m_2, ..., m_n) $, (and $\sum_{\mathbf{m}} $ is the sum over all possible $\mathbf{m}$), $\ket{j_s, \mathbf{m}}_s  = \bigotimes_{i=1}^{n} \ket{j_s, m_i}_s $, and  $\mathcal{D}_{\mathbf{m}, -j_s}^{(j_s)}  = \Pi_{i=1}^{n} d_{m_i, -j_s}^{(j_s)} \left( \phi_i \right)$. We will further use $\boldsymbol{\theta} = (\theta_1, \theta_2, ..., \theta_n) $.
After the off-resonance interaction, the spins-cavity state is:
\begin{equation}
    \ket{\Psi (t_R + t_I) }_{sc} = \sum_{\mathbf{m}} \mathcal{D}_{\mathbf{m}, -j_s}^{(j_s)} \Ket{j_s,\mathbf{m}}_s \Ket{\alpha_0 e^{- i \boldsymbol{\theta} \cdot \mathbf{m} }}_c
\end{equation}
where $\boldsymbol{\theta} \cdot \mathbf{m} = \sum_{i=1}^n \theta_i m_i$. 

The found results will also hold for half-integer spins with $\alpha_0 = \frac{1+i}{\sqrt{2} } \left| \alpha_0 \right| $. Taking the projective measurement, the UPM density matrix of the spins, when the outcome $\pm$ is obtained, are:
\begin{equation}
    \rho^{\text{U}}_{{\pm}_s} = \sum_{\mathbf{m} ,\mathbf{m'} } \mathcal{D}_{\mathbf{m}, -j_s}^{(j_s)} \mathcal{D}_{\mathbf{m'}, -j_s}^{(j_s)}  \mathcal{I}_{\mathbf{m} ,\mathbf{m'}}^\pm \Ket{j_s,\mathbf{m}}  \bra{j_s,\mathbf{m'}}_s 
\end{equation}
where 
\begin{align}
    \mathcal{I}_{\mathbf{m} ,\mathbf{m'}}^+ &:= \int_{0}^{\infty} dx \langle x | e^{- i \boldsymbol{\theta} \cdot \mathbf{m} } \alpha_0\rangle_c \langle e^{- i \boldsymbol{\theta} \cdot \mathbf{m'} } \alpha_0| x\rangle_c \nonumber \\
    &=  \frac{1}{\sqrt{\pi}} \int_{0}^{\infty} \, dx \,  \text{exp}\left[ - \frac{1}{2}  \left( \left(x - \sqrt{2} \alpha_0 \cos ( \boldsymbol{\theta} \cdot \mathbf{m} )  \right)^2 + \left(x - \sqrt{2} \alpha_0 \cos ( \boldsymbol{\theta} \cdot \mathbf{m'} )  \right)^2 \right) + i x \sqrt{2} \alpha_0  ( \sin ( \boldsymbol{\theta} \cdot \mathbf{m} )   -  \sin ( \boldsymbol{\theta} \cdot \mathbf{m'} )  )  \right]   \nonumber \\
    &=  \frac{1}{2} \left(1+\text{erf}\left(\frac{\alpha_0  (i (\sin (\boldsymbol{\theta} \cdot \mathbf{m}  )-\sin (\boldsymbol{\theta} \cdot \mathbf{m'} ))+\cos (\boldsymbol{\theta} \cdot \mathbf{m} )+\cos (\boldsymbol{\theta} \cdot \mathbf{m'} ))}{\sqrt{2}}\right)\right)  \nonumber \\
    & \hspace{2.5cm} \exp \left(\frac{1}{2} \alpha_0 ^2 (2 \cos (\boldsymbol{\theta} \cdot \mathbf{m} -\boldsymbol{\theta} \cdot \mathbf{m'}  )+i (2 \sin (\boldsymbol{\theta} \cdot \mathbf{m} -\boldsymbol{\theta} \cdot \mathbf{m'} )+\sin (2 \boldsymbol{\theta} \cdot \mathbf{m}  ) -\sin (2 \boldsymbol{\theta} \cdot \mathbf{m'} )+2 i))\right),
\end{align}
and similarly for the $-$ measurement, with the difference in the integration limits. 

With the second rotation every $\ket{j_s, m_i}_s$ will be rotated to $ \sum_{n_i = -j_s}^{j_s} d_{n_i,m_i}^{(j_s)} \left( \phi_i \right) \Ket{j_s,n_i}_s$. Hence, the UPM state will be rotated to:
\begin{equation}
    \rho^{\text{U}}_{{\pm}_s}(t_R) = \sum_{\mathbf{m} ,\mathbf{m'}, \mathbf{n} ,\mathbf{n'} } \mathcal{D}_{\mathbf{m}, -j_s}^{(j_s)} \mathcal{D}_{\mathbf{m'}, -j_s}^{(j_s)} \mathcal{D}_{\mathbf{n}, \mathbf{m}}^{(j_s)} \mathcal{D}_{\mathbf{n'}, \mathbf{m'}}^{(j_s)}  \mathcal{I}_{\mathbf{m} ,\mathbf{m'}}^\pm \Ket{j_s,\mathbf{n}}  \bra{j_s,\mathbf{n'}}_s,
\end{equation}
Taking the second projective measurement, when the outcome $\pm$ is obtained for the first measurement and the outcome $+$ is obtained for the second measurement, the UPM spin states are given by, 
\begin{equation}
    \rho^{\text{U}}_{{\pm +_s}} = \sum_{\mathbf{m} ,\mathbf{m'}, \mathbf{n} ,\mathbf{n'} } \mathcal{D}_{\mathbf{m}, -j_s}^{(j_s)} \mathcal{D}_{\mathbf{m'}, -j_s}^{(j_s)} \mathcal{D}_{\mathbf{n} , \mathbf{m}}^{(j_s)} \mathcal{D}_{\mathbf{n'}, \mathbf{m'}}^{(j_s)}  \mathcal{I}_{\mathbf{m} ,\mathbf{m'}}^\pm \mathcal{I}_{\mathbf{n} ,\mathbf{n'}}^+ \Ket{j_s,\mathbf{n}}  \bra{j_s,\mathbf{n'}}_s.
\end{equation}
Tracing out the spin Hilbert space,
\begin{equation}
    P_{1,2} (\pm +) = \sum_{\mathbf{m} ,\mathbf{m'}, \mathbf{n} } \mathcal{D}_{\mathbf{m}, -j_s}^{(j_s)} \mathcal{D}_{\mathbf{m'}, -j_s}^{(j_s)} \mathcal{D}_{\mathbf{n} , \mathbf{m}}^{(j_s)} \mathcal{D}_{\mathbf{n}, \mathbf{m'}}^{(j_s)}  \mathcal{I}_{\mathbf{m} ,\mathbf{m'}}^\pm \mathcal{I}_{\mathbf{n} ,\mathbf{n}}^+.
\end{equation}
Note that the diagonal terms are
\begin{equation}
    \mathcal{I}_{\mathbf{n} ,\mathbf{n}}^+ = \frac{1}{\sqrt{\pi}} \int_{0}^{\infty} \,  dx \, \text{exp}\left[ - \left(x - \sqrt{2} \alpha_0 \cos ( \boldsymbol{\theta} \cdot \mathbf{n} )  \right)^2  \right] = \frac{1}{2} \left( 1 + \text{erf}(\sqrt{2} \alpha_0 \cos ( \boldsymbol{\theta} \cdot \mathbf{n} ))\right)
\end{equation}
Hence, 
\begin{equation}
    P_{2} (\pm) = \sum_{\mathbf{m} ,\mathbf{m'}, \mathbf{n} } \mathcal{D}_{\mathbf{m}, -j_s}^{(j_s)} \mathcal{D}_{\mathbf{m'}, -j_s}^{(j_s)} \mathcal{D}_{\mathbf{n} , \mathbf{m}}^{(j_s)} \mathcal{D}_{\mathbf{n}, \mathbf{m'}}^{(j_s)}  \frac{1}{2} \left( 1 \pm \text{erf}(\sqrt{2} \alpha_0 \cos ( \boldsymbol{\theta} \cdot \mathbf{n} ))\right)
\end{equation}

In this case, it is not possible to cancel this noise by taking large enough $\alpha_0$, because, depending on the $\sigma_g$, the value of $\cos ( \boldsymbol{\theta} \cdot \mathbf{m} )$ could approach a null value, and similarly the $\sin$ in the off-diagonal part. Given the derived probabilities, the violations are numerically computed.

\section{Inhomogeneity from Physical Implementation \label{app:physical}}

The solution of the electric field -- and hence of the modes considered --  is needed to describe a specific physical implementation. By rewriting Maxwell's equations in wave form, in vacuum the electric field is governed by the Helmholtz equation $ \nabla^2 \mathbf{E} + k_0^2 \mathbf{E} = 0$, where $k_0 = \omega/c$, $\mathbf{E} = (E_x,  E_y ,  E_z)^T$, and $\mathbf{E}(t) = \mathbf{E} e^{- i \omega t}$ (and similarly for $\mathbf{B}$). The type of cavity of Fig.~\ref{fig:diagram_exp_SetUp} allows two families of solutions, known as Transverse Electric (TE) and Transverse Magnetic (TM) modes, having $E_z = 0$ or $B_z = 0$, respectively. For TM modes, in this work considered, all components of the electromagnetic field are uniquely determined by the $z$-components. In fact, let $\textbf{E} = \textbf{E}_{\perp} + E_z \hat{\mathbf{z}}$ where $\textbf{E}_{\perp} = (E_x, E_y)$ are the transverse components and and $E_z$ the propagating component. In this form, Maxwell's equation in the vacuum implies that and 
\begin{equation}
\label{eq:derivation_TM_Identity}
     (\nabla \times \mathbf{E})_z = 0 \implies \partial_x E_y = \partial_y E_x \implies \nabla_{\perp} \times \textbf{E}_{\perp} = 0 \implies  \textbf{E}_{\perp} = \nabla_{\perp} \psi \implies \nabla_{\perp} ^2 \psi = - \partial_z E_z
\end{equation}
where the first identity follows from $B_z = 0$, $\psi$ is a scalar potential, and the last identity follows from the other Maxwell's equation, which reads $\nabla \mathbf{E} = \nabla_{\perp} \textbf{E}_{\perp} + \partial_z E_z = 0 $. This last Maxwell's equation can be approached by separation of variables, $E_z = \phi(x,y) Z(z)$, and the mode expansion leads to the eigenvalues solution
\begin{equation}
\label{eq:derivation_TM_Identity}
    \nabla_{\perp}^2 \phi(x,y) + k_{\perp}^2 \phi(x,y) = 0  \;\;\; \text{and} \;\;\; \partial_z^2 Z(z) + k_{z}^2 Z(z) = 0 \;\;\; \text{with} \;\;\; k_{z}^2  = k_0^2 - k_{\perp}^2 
\end{equation}
where the latter, i.e. the dissipation relation, constrains the wavevector on the $z$-axis given the frequency of the experiment ($\omega$) and the eigenvalues from the $xy$ boundary conditions. Often, as we shall see, this leads to complex values of $k_z$ and hence a decay behaviour in the $z$ direction, known as evanescent TM modes.

The final identity of Eq.~(\ref{eq:derivation_TM_Identity}), for $\psi = \phi(x,y) f(z)$, reads
\begin{equation}
\label{eq:electric_field_TM}
    - k_{\perp}^2 \phi(x,y) f(z) = -  \phi(x,y)  \partial_z Z(z) \implies \psi = \frac{1}{ k_{\perp}^2 } \partial_z E_z \implies \textbf{E}_{\perp} = \frac{1}{ k_{\perp}^2 }  \nabla_{\perp} (\partial_z E_z )
\end{equation}
explicitly, showing the relationship between the transverse components and the $z$-components, known as TM identity. 

Let us seek the particular solution under the boundary conditions imposed by the conducting box, i.e. at $x= \{- L_x/2, L_x/2 \}$ and  $y= \{-L_y/2, L_y/2 \}$, $E_z = 0 \; \forall z < L_z$ and $E_x, E_y =0 $ for $z = z_0$. The former implies that the solutions for the transverse modes inside the box are 
\begin{equation}
    E_z^{(i)}(x,y,z, t) = E_0 \phi_{mn}(x,y)e^{i k_z^{(i)} z} e^{-i \omega t} = E_0 \cos \left(  \frac{ m \pi x}{L_x} \right) \cos \left(  \frac{ n \pi y}{L_y} \right) \left(A_z \cos(k_z^{(i)} z) + B_z \sin(k_z^{(i)} z) \right) e^{-i \omega t} 
\end{equation}
where $E_0$ is the vacuum electric field, $m$ and $n$ are the labels of the harmonics $\phi_{mn}$ and $(i)$, represents the fact that these are quantities inside the box, and $|A_z|^2 +  |B_z|^2 = 1$ (as the normalization is given by $E_0$). From $E_x, E_y =0 $ for $x = z_0$, it is possible to show that $B_z = 0 \implies A_z = 1$, as from Eq.~(\ref{eq:electric_field_TM}), the other components can be computed
\begin{equation}
    \textbf{E}_{\perp}^{(i)} (x,y,z, t)  = -  E_0 \frac{k_z^{(i)}}{ k_{\perp}^2 } \sin(k_z^{(i)} z) \begin{pmatrix}
        \partial_x \phi_{mn}(x,y) \\
        \partial_y \phi_{mn}(x,y) 
    \end{pmatrix} e^{-i \omega t} 
\end{equation}
which are indeed zero at $z_0 = 0$. From the continuity at $z = L_z$, we require $\mathbf{E}^{(i)} (x,y, L_z, t) =  \mathbf{E}^{(o)} (x,y, L_z, t) \; \forall \; t,x,y$. The outside $z$ component of the field can be taken to be 
\begin{equation}
     E_z^{(o)}(x,y, z, t) = E_0 \cos(k_z^{(i)} L_z) \phi_{mn}(x,y)e^{ \pm i (z - L_z) k_z^{(o)} } e^{-i \omega t} \;,
\end{equation}
such that $k_{\perp}^{(o)} = k_{\perp}^{(i)} =k_{\perp}$, as $\phi_{mn}(x,y)$ do not change outside. Additionally, from the other components derived (from Eq.~\ref{eq:electric_field_TM}) evaluated at $L_z$, it is possible to get the ratio
\begin{equation}
    1 = \frac{E_x^{(i)} (x,y, L_z, t) }{E_x^{(o)} (x,y, L_z, t) } = \left. \frac{ \partial_z E_z^{(i)}}{\cos ( k^{(i)}_z L_z )   \partial_z E_z^{(o)} } \right|_{z = L_z } = \frac{ - k^{(i)}_z \sin ( k^{(i)}_z L_z ) }{ i  k^{(o)}_z \cos ( k^{(i)}_z L_z ) } \implies k^{(o)}_z = - i k^{(i)}_z \tan ( k^{(i)}_z L_z )
\end{equation}
defining $\beta = k^{(i)}_z \tan ( k^{(i)}_z L_z )$,  which implies an exponential decay outside the resonators, i.e. the field is
\begin{equation}
     E_z^{(o)}(x,y, z, t) = E_0 \cos(k_z^{(i)} L_z) \phi_{mn}(x,y)e^{ - \beta (z - L_z) } e^{-i \omega t} \;.
\end{equation}


The chosen parameters of the resonator are $L_x \sim 2 \cdot 10^{-2}$ m, $L_y \sim 8 \cdot 10^{-6}$ m, and $L_z \sim 2 \cdot 10^{-7}$, such that $V \sim 3.2 \cdot 10^{-14}\text{ m}^3$. Taking into account the resonant $\omega$, fundamental to to $L_x$, i.e. $m=n=1$, these values imply that inside the cavity, $k_z^{(i)} = \sqrt{\frac{\omega^2}{c^2} - \frac{\pi^2}{L_x^2} - \frac{\pi^2}{L_y^2} } \sim i \frac{\pi}{L_y} \sim i 4 \cdot  10^{5}$, which, for $z < L_z$ is negligible. However, outside the resonator, $k_z^{(o)} \sim i 3 \cdot 10^{-4}$. For a Rygdberg atom placed at $z\approx 50 \; \mu$m, this implies a decay of a factor of $e^{-1.5}\sim 0.2$.




Then, the coupling as function of the qubit's position inside the resonator is approximately given by (omitting the $z$-dependency for the previous argument)
\begin{equation}
g (x,y) =  \frac{\mathbf{E}^{(i)}(x,y) \cdot \mathbf{d}}{\hbar} = \frac{d E_z^{(i)}}{\hbar }= \frac{E_0 d}{ \hbar} \cos\left( \frac{x \pi}{L_x}\right) \cos\left( \frac{y \pi}{L_y}\right)  
\end{equation}

For an arbitrary qubit in the configuration $x = x_k = r_q \cdot k $ and $y = 0$ for $k \in [ -N/2, N/2-1 ]$, implying an ensemble of $N$ qubits. One then finds that 
\begin{equation}
    g_k = g   \cos\left( \frac{\pi r_q}{L_x} k \right) \approx g \left(1 -  \frac{\pi^2  r_q^2}{2 L_x^2} k^2 \right)  = g \left(1 -  \frac{1}{2} \alpha^2 k^2 \right) 
\end{equation}
where we have defined that  $g= E_0 d/(\hbar)$, which leads to Eq.~(\ref{eq:physical_coupling}), $L_x = L$, $\alpha = \pi r_q/ L$ and the Taylor expansion is taken in the limit:$1 \gg \alpha N \gg \alpha$ (i.e. we have assumed that the the qubits are placed along the long side of the resonator (Fig.~\ref{fig:diagram_exp_SetUp}), but only within a small fraction of the long side, i.e., $r_q N/L \ll 1$). One can compute the  mean coupling and its standard deviation as
 \begin{align}
    \braket{g} &= \frac{1}{N} \sum_{k = -N/2}^{N/2-1} g_k = g \left(1 - \frac{\alpha^2 }{2 N} \sum_{k = -N/2}^{N/2-1} k^2 \right) = g \left(1 - \frac{\alpha^2}{24} \left(N^2 + 2 \right) \right)  \approx g \left(1 - \frac{\alpha^2 N^2}{24} \right),
\end{align}
and
\begin{align}
    (\sigma_g)^2 &= \frac{1}{N} \sum_{k = -N/2}^{N/2-1} (g_k)^2 - (\braket{g})^2 = - (\braket{g})^2 + g^2 \left(1 -  \frac{ \alpha^2}{ N}  \sum_{k = -N/2}^{N/2-1} k^2 + \frac{ \alpha^4}{4 N} \sum_{k = -N/2}^{N/2-1} k^4 \right) \nonumber \\
    &= \frac{g^2 \alpha^4}{720} \left( - 11 + 10 N^2 +  N^4 \right) \approx \frac{g^2 \alpha^4 N^4}{720},
\end{align} 
where the last approximations are for large $N$. Then, $\sigma_g  \lesssim \mathcal{O} (g /(10 N))$ implies the following:
\begin{equation}
    \frac{\sigma_g}{\braket{g}} \approx \frac{\alpha^2 N^2}{12 \sqrt{5}\left(1 - \alpha^2 N^2/24\right)} \approx \frac{\alpha^2 N^2}{12 \sqrt{5}} = \frac{\pi^2 r_q^2 N^2}{12 \sqrt{5} L^2} \lesssim \frac{1}{10 N} \implies N_{\text{NDC}}  \lesssim 0.6 \left( \frac{L}{r_q} \right)^{2/3}.
\end{equation}


\begin{thebibliography}{1}

\bibitem{Athalye_2011} V. Athalye, S. S. Roy, and T. S. Mahesh, \emph{Investigation of the Leggett-Garg Inequality for Precessing Nuclear Spins}, \href{https://journals.aps.org/prl/abstract/10.1103/PhysRevLett.107.130402}{Phys. Rev. Lett. {\bf 107}, 130402 (2011)}.

\bibitem{knee2012violation} G. C. Knee, S. Simmons, \textit{et al.} \emph{Violation of a Leggett--Garg inequality with ideal non-invasive measurements}, \href{https://www.nature.com/articles/ncomms1614}
{Nature Communications \textbf{3}, 606 (2012)}.




\bibitem{newton} A. Vaartjes, M. Nurizzo, \textit{et al.}, \emph{Certifying the quantumness of a nuclear spin qudit through its uniform precession}, \href{https://doi.org/10.1016/j.newton.2025.100017}{Newton \textbf{1}, 100017 (2025)}.





\bibitem{lgi1} A. J. Leggett and A. Garg, \emph{Quantum mechanics versus macroscopic realism: Is the flux there when nobody looks?}, \href{https://doi.org/10.1103/PhysRevLett.54.857}{Phys. Rev. Lett. \textbf{54}, 857 (1985)}.

\bibitem{leggett02} A. J. Leggett, \emph{Testing the limits of quantum mechanics: motivation, state of play, prospects}, \href{https://doi.org/10.1088/0953-8984/14/15/201}{J. Phys. Condens. Matter \textbf{14}, R415 (2002)}.

\bibitem{lgi2} A. J. Leggett, \emph{Realism and the physical world}, \href{https://doi.org/10.1088/0034-4885/71/2/022001}{Rep. Prog. Phys. \textbf{71}, 022001 (2008)}.

\bibitem{qlgi1} C. Emary, N. Lambert, and F. Nori, \emph{Leggett–Garg inequalities}, \href{https://doi.org/10.1088/0034-4885/77/1/016001}{Rep. Prog. Phys. \textbf{77},	016001 (2014)}.

\bibitem{nsit1} J. Kofler and C. Brukner, \emph{Condition for macroscopic realism beyond the Leggett-Garg inequalities}, \href{https://doi.org/10.1103/PhysRevA.87.052115}{Phys. Rev. A \textbf{87}, 052115 (2013)}.

\bibitem{nsit2}	G. Schild and C. Emary, \emph{Maximum violations of the quantum-witness equality}, \href{https://doi.org/10.1103/PhysRevA.92.032101}{Phys. Rev. A \textbf{92}, 032101 (2015)}.

\bibitem{nsit3} G. C. Knee, K. Kakuyanagi \textit{et al.}, \emph{A strict experimental test of macroscopic realism in a superconducting flux qubit}, \href{https://doi.org/10.1038/ncomms13253}{Nature Communications \textbf{7}, 13253 (2016)}.

\bibitem{largemass} D. Das, D. Home, H. Ulbricht, and S. Bose, \emph{Mass-Independent Scheme to Test the Quantumness of a Massive Object}, \href{https://doi.org/10.1103/PhysRevLett.132.030202}{Phys. Rev. Lett. \textbf{132}, 030202 (2024)}.

\bibitem{gravity_measurement} F. Hanif, D. Das \textit{et al.}, \emph{Testing Whether Gravity Acts as a Quantum Entity When Measured}, \href{https://doi.org/10.1103/PhysRevLett.133.180201}{Phys. Rev. Lett. \textbf{133}, 180201 (2024)}.



\bibitem{computing1} J. H. Wesenberg, A. Ardavan \textit{et al.}, \emph{Quantum Computing with an Electron Spin Ensemble}, \href{https://doi.org/10.1103/PhysRevLett.103.070502}{Phys. Rev. Lett. \textbf{103}, 070502 (2009)}.

\bibitem{memory1} Y. Kubo, F. R. Ong \textit{et al.}, \emph{Strong Coupling of a Spin Ensemble to a Superconducting Resonator}, \href{https://doi.org/10.1103/PhysRevLett.105.140502}{Phys. Rev. Lett. {\bf 105}, 140502 (2010)}.

\bibitem{Chen2018} J. Chen, A. A. Zadorozhko, and D. Konstantinov, 
\emph{Strong coupling of a two-dimensional electron ensemble to a single-mode cavity resonator}, 
\href{https://doi.org/10.1103/PhysRevB.98.235418}{Phys. Rev. B \textbf{98}, 235418 (2018)}. 

\bibitem{computing2} Y. Ping, E. M. Gauger, and S. C. Benjamin, \emph{Measurement-based quantum computing with a spin ensemble coupled to a stripline cavity}, \href{https://iopscience.iop.org/article/10.1088/1367-2630/14/1/013030}{New J. Phys. {\bf 14}, 013030 (2012)}.

\bibitem{qubit_ensemble_computing_1} N. Mohseni, M. Narozniak \textit{et al.}, \emph{Error suppression in adiabatic quantum computing with qubit ensembles}, \href{https://doi.org/10.1038/s41534-021-00405-2}{npj Quantum Inf \textbf{7}, 71 (2021)}. 

\bibitem{sensing1} F. Troiani, A. Ghirri, M. G. A. Paris, C. Bonizzoni, and M. Affronte, \emph{Towards quantum sensing with molecular spins}, \href{https://doi.org/10.1016/j.jmmm.2019.165534}{Journal of Magnetism and Magnetic Materials {\bf 491},  165534 (2019)}.

\bibitem{sensing2} H. Wu, S. Yang \textit{et al.}, \emph{Enhanced quantum sensing with room-temperature solid-state masers, }\href{https://www.science.org/doi/10.1126/sciadv.ade1613}{Science Advances {\bf 8},  48 (2022)}.

\bibitem{metro} M. Schaffry, E. M. Gauger \textit{et al.}, \emph{Quantum metrology with molecular ensembles}, \href{https://doi.org/10.1103/PhysRevA.82.042114}{Phys. Rev. A \textbf{82}, 042114 (2010)}.

\bibitem{morton_memory_1} H. Wu, R. E. George \textit{et al.}, \emph{Storage of Multiple Coherent Microwave Excitations in an Electron Spin Ensemble}, \href{https://doi.org/10.1103/PhysRevLett.105.140503}{Phys. Rev. Lett. \textbf{105}, 140503 (2010)}.


\bibitem{memory2}  D. I. Schuster, A. P. Sears \textit{et al.}, \emph{High-Cooperativity Coupling of Electron-Spin Ensembles to Superconducting Cavities}, \href{https://doi.org/10.1103/PhysRevLett.105.140501}{Phys. Rev. Lett. \textbf{105}, 140501 (2010)}.

\bibitem{mem} Y. Kubo, C. Grezes \textit{et al.}, \emph{Hybrid Quantum Circuit with a Superconducting Qubit Coupled to a Spin Ensemble}, \href{https://doi.org/10.1103/PhysRevLett.107.220501}{Phys. Rev. Lett. \textbf{107}, 220501 (2011)}.

\bibitem{mem1} X. Zhu, S. Saito \textit{et al.}, \emph{Coherent coupling of a superconducting flux qubit to an electron spin ensemble in diamond}, \href{https://doi.org/10.1038/nature10462}{Nature \textbf{478}, 221-224 (2011)}.

\bibitem{memory3} R. Amsüss, Ch. Koller \textit{et al.}, \emph{Cavity QED with Magnetically Coupled Collective Spin States}, \href{https://doi.org/10.1103/PhysRevLett.107.060502}{Phys. Rev. Lett. \textbf{107}, 060502 (2011)}.

\bibitem{mem4} J. H. Wesenberg, Z. Kurucz, and K. Molmer, \emph{Dynamics of the collective modes of an inhomogeneous spin ensemble in a cavity}, \href{https://doi.org/10.1103/PhysRevA.83.023826}{Phys. Rev. A \textbf{83}, 023826 (2011)}.

\bibitem{mem2}  S. Saito, X. Zhu, \textit{et al.}, \emph{Towards Realizing a Quantum Memory for a Superconducting Qubit: Storage and Retrieval of Quantum States}, \href{https://doi.org/10.1103/PhysRevLett.111.107008}{Phys. Rev. Lett. \textbf{111}, 107008 (2013)}.

\bibitem{memory4} V. Ranjan, G. de Lange, \textit{et al.}, \emph{Probing Dynamics of an Electron-Spin Ensemble via a Superconducting Resonator}, \href{https://doi.org/10.1103/PhysRevLett.110.067004}{Phys. Rev. Lett. \textbf{110}, 067004 (2013)}.

\bibitem{memory5} S. Probst, H. Rotzinger \textit{et al.}, \emph{Anisotropic Rare-Earth Spin Ensemble Strongly Coupled to a Superconducting Resonator}, \href{https://doi.org/10.1103/PhysRevLett.110.157001}{Phys. Rev. Lett. \textbf{110}, 157001 (2013)}.

\bibitem{mem3} B. Julsgaard, and K. Molmer, \emph{Fundamental limitations in spin-ensemble quantum memories for cavity fields}, \href{https://doi.org/10.1103/PhysRevA.88.062324}{Phys. Rev. A  \textbf{88}, 062324 (2013)}.

\bibitem{mmry} C. Grezes, B. Julsgaard \textit{et al.}, \emph{Multimode Storage and Retrieval of Microwave Fields in a Spin Ensemble}, \href{https://doi.org/10.1103/PhysRevX.4.021049}{Phys. Rev. X {\bf 4}, 021049 (2014)}.


\bibitem{memory6} Y. Tabuchi, S. Ishino \textit{et al.}, \emph{Hybridizing Ferromagnetic Magnons and Microwave Photons in the Quantum Limit}, \href{https://doi.org/10.1103/PhysRevLett.113.083603}{Phys. Rev. Lett. \textbf{113}, 083603 (2014)}.

\bibitem{memory7} C. Grezes, B. Julsgaard \textit{et al.}, \emph{Storage and retrieval of microwave fields at the single-photon level in a spin ensemble}, \href{https://doi.org/10.1103/PhysRevA.92.020301}{Phys. Rev. A \textbf{92}, 020301(R) (2015)}.

\bibitem{memory8} C. Grezes, Y. Kubo \textit{et al.}, \emph{Towards a spin-ensemble quantum memory for superconducting qubits}, \href{https://doi.org/10.1016/j.crhy.2016.07.006}{Comptes Rendus Physique \textbf{17}, 693 (2016)}.

\bibitem{memory9} J. J. L. Morton, and  P. Bertet, \emph{Storing quantum information in spins and high-sensitivity ESR}, \href{https://doi.org/10.1016/j.jmr.2017.11.015}{Journal of Magnetic Resonance \textbf{287}, 128 (2018)}. 

\bibitem{memory10} J. O'Sullivan, O. W. Kennedy \textit{et al.}, \emph{Random-Access Quantum Memory Using Chirped Pulse Phase Encoding}, \href{https://doi.org/10.1103/PhysRevX.12.041014}{Phys. Rev. X {\bf 12}, 041014 (2022)}.

\bibitem{memory11} M. H. Appel, A. Ghorbal \textit{et al.}, \emph{Many-body quantum register for a spin qubit}, \href{https://doi.org/10.1038/s41567-024-02746-z}{Nat. Phys. \textbf{21}, 368-373 (2025)}.

\bibitem{comm} S. Simmons, R. Brown \textit{et al.}, \emph{Entanglement in a solid-state spin ensemble}, \href{https://doi.org/10.1038/nature09696}{Nature \textbf{470}, 69 (2011)}.

\bibitem{comm1} M. Zhong, M. Hedges \textit{et al.}, \emph{Optically addressable nuclear spins in a solid with a six-hour coherence time}, \href{https://doi.org/10.1038/nature14025}{Nature {\bf 517}, 177 (2015)}.

\bibitem{comm2} A. Bourassa, C. P. Anderson \textit{et al.}, \emph{Entanglement and control of single nuclear spins in isotopically engineered silicon carbide}, \href{https://doi.org/10.1038/s41563-020-00802-6}{Nat. Mater. \textbf{19}, 1319–1325 (2020)}.





\bibitem{HP_transformation} T. Holstein, and H. Primakoff,
\emph{Field Dependence of the Intrinsic Domain Magnetization of a Ferromagnet}, 
\href{https://doi.org/10.1103/PhysRev.58.1098}{Phys. Rev. \textbf{58}, 1098 (1940)}.

\bibitem{sadana2022testing} S. Sadana, L. Maccone, and U. Sinha, \emph{Testing quantum foundations with quantum computers}, \href{https://doi.org/10.1103/PhysRevResearch.4.L022001}{Phys. Rev. Research \textbf{4}, L022001 (2022)}.

\bibitem{joarder2022loophole} K. Joarder, D. Saha, D. Home, and U. Sinha, \emph{Loophole-Free Interferometric Test of Macrorealism Using Heralded Single Photons}, \href{https://doi.org/10.1103/PRXQuantum.3.010307}{PRX Quantum \textbf{3}, 010307 (2022)}.

\bibitem{sadana2023noise} S. Sadana, L. Maccone, and U. Sinha, \emph{Noise analysis for the Sorkin and Peres tests performed on a quantum computer}, \href{https://doi.org/10.1007/s11128-023-04072-4}{Quantum Inf Process \textbf{22}, 317 (2023)}.

\bibitem{nath2024single} P. P. Nath, D. Saha, D. Home, and U. Sinha, \emph{Single-System-Based Generation of Certified Randomness Using Leggett-Garg Inequality}, \href{https://doi.org/10.1103/PhysRevLett.133.020802}{Phys. Rev. Lett. \textbf{133}, 020802 (2024)}.

\bibitem{lsnew1} N. Gisin, and A. Peres, \emph{Maximal violation of Bell's inequality for arbitrarily large spin}, \href{https://doi.org/10.1016/0375-9601(92)90949-M}{Phys. Lett. A \textbf{162}, 15 (1992)}.

\bibitem{lsnew2} A. Peres, \emph{Finite violation of a Bell inequality for arbitrarily large spin}, \href{https://doi.org/10.1103/PhysRevA.46.4413}{Phys. Rev. A \textbf{46}, 4413 (1992)}.

\bibitem{lsnew3} D. Home, and A. S. Majumdar, \emph{Incompatibility between quantum mechanics and classical realism in the ``strong'' macroscopic limit}, \href{https://doi.org/10.1103/PhysRevA.52.4959}{Phys. Rev. A \textbf{52}, 4959 (1995)}.

\bibitem{lsnew4} A. Cabello, \emph{Bell's inequality for $n$ spin-$s$ particles}, \href{https://doi.org/10.1103/PhysRevA.65.062105}{Phys. Rev. A \textbf{65}, 062105 (2002)}.

\bibitem{large_spin1} J. Kofler, and C. Brukner, \emph{Classical World Arising out of Quantum Physics under the Restriction of Coarse-Grained Measurements}, \href{https://journals.aps.org/prl/abstract/10.1103/PhysRevLett.99.180403}{Phys. Rev. Lett. {\bf 99}, 180403 (2007)}.


\bibitem{large_spin2} J. Kofler, and C. Brukner, \emph{Conditions for Quantum Violation of Macroscopic Realism}, \href{https://journals.aps.org/prl/abstract/10.1103/PhysRevLett.101.090403}{Phys. Rev. Lett. {\bf 101}, 090403 (2008)}.

\bibitem{lsnew5} C. Budroni, T. Moroder, M. Kleinmann, and O. Guhne, \emph{Bounding Temporal Quantum Correlations}, \href{https://doi.org/10.1103/PhysRevLett.111.020403}{Phys. Rev. Lett. \textbf{111}, 020403 (2013)}.

\bibitem{large_spin3} H. Jeong, Y. Lim, and M. S. Kim, \emph{Coarsening Measurement References and the Quantum-to-Classical Transition}, \href{https://journals.aps.org/prl/abstract/10.1103/PhysRevLett.112.010402}{Phys. Rev. Lett. {\bf 112}, 010402 (2014)}.


\bibitem{large_spin4} C. Budroni, and C. Emary, \emph{Temporal Quantum Correlations and Leggett-Garg Inequalities in Multilevel Systems}, \href{https://journals.aps.org/prl/abstract/10.1103/PhysRevLett.113.050401}{Phys. Rev. Lett. {\bf 113}, 050401 (2014)}.

\bibitem{large_spin5} S. Mal, and A. S. Majumdar, \emph{Optimal violation of the Leggett-Garg inequality for arbitrary spin and emergence of classicality through unsharp measurements}, \href{https://www.sciencedirect.com/science/article/pii/S0375960116302122}{Phys. Lett. A {\bf 380}, 2265 (2016)}.

\bibitem{large_spin6} S. Mal, D. Das, and D. Home, \emph{Quantum mechanical violation of macrorealism for large spin and its robustness against coarse-grained measurements}, \href{https://journals.aps.org/pra/abstract/10.1103/PhysRevA.94.062117}{Phys. Rev. A {\bf 94}, 062117 (2016)}.

\bibitem{large_spin7} S. Mukherjee, A. Rudra, D. Das, S. Mal, and D. Home, \emph{Persistence of quantum violation of macrorealism for large spins even under coarsening of measurement times}, \href{https://doi.org/10.1103/PhysRevA.100.042114}{Phys. Rev. A \textbf{100}, 042114 (2019)}.





\bibitem{s9} S. Hill, R. S. Edwards, N. Aliaga-Alcalde, and G. Christou, \emph{Quantum coherence in an exchange-coupled dimer of single-molecule magnets}, \href{https://www.science.org/doi/10.1126/science.1090082}{Science {\bf 302}, 1015 (2003)}.

\bibitem{s10-1} E. del Barco, N. Vernier \textit{et al.}, \emph{Quantum coherence in Fe8 molecular nanomagnets}, \href{https://iopscience.iop.org/article/10.1209/epl/i1999-00450-8}{EPL {\bf 47}, 722 (1999)}.

\bibitem{s10-2} E. del Barco, J. M. Hernandez \textit{et al.}, \emph{High-frequency resonant experiments in Fe$_8$ molecular clusters}, \href{https://doi.org/10.1103/PhysRevB.62.3018}{Phys. Rev. B {\bf 62}, 3018 (2000)}.

\bibitem{s150-1} D. D. Awschalom, D. P. DiVincenzo, and J. F. Smyth, \emph{Macroscopic quantum effects in nanometer-scale magnets}, \href{https://www.science.org/doi/10.1126/science.258.5081.414}{Science \textbf{258}, 414 (1992)}.

\bibitem{s150-2} D. D. Awschalom, J. F. Smyth, G. Grinstein, D. P. DiVincenzo, and D. Loss, \emph{Macroscopic quantum tunneling in magnetic proteins}, \href{https://doi.org/10.1103/PhysRevLett.68.3092}{Phys.
Rev. Lett. \textbf{68}, 3092 (1992).}

\bibitem{s150-3} S. Gider, D. Awschalom, T. Douglas, S.  Mann and M. Charala, \emph{Classical and quantum magnetic phenomena in natural and artificial ferritin proteins}, \href{https://www.science.org/doi/10.1126/science.7701343}{Science {\bf 268}, 77 (1995)}.

\bibitem{BEC} H. W. Lau, Z. Dutton, T. Wang, and C. Simon, \emph{Proposal for the Creation and Optical Detection of Spin Cat States in Bose-Einstein Condensates}, \href{https://doi.org/10.1103/PhysRevLett.113.090401}{Phys. Rev. Lett. {\bf 113}, 090401 (2014)}.

\bibitem{Nucleus} P. Gupta, A. Vaartjes, X. Yu, A. Morello, and B. C. Sanders, \emph{Robust Macroscopic Schr\"{o}dinger's Cat on a Nucleus}, \href{https://doi.org/10.1103/PhysRevResearch.6.013101}{Phys. Rev. Research \textbf{6}, 013101 (2024)}.


\bibitem{qi2} C. M. Ramsey, E. del Barco \textit{et al.}, \emph{Quantum interference of tunnel trajectories between states of different spin length in a dimeric molecular nanomagnet}, \href{https://doi.org/10.1038/nphys886}{Nature Phys. {\bf 4}, 277 (2008)}.



\bibitem{Xiang2013} Z.-L. Xiang, S. Ashhab, J. Q. You, and F. Nori, 
\emph{Hybrid quantum circuits: Superconducting circuits interacting with other quantum systems}, 
\href{https://doi.org/10.1103/RevModPhys.85.623}{Rev. Mod. Phys. \textbf{85}, 623–653 (2013)}.

\bibitem{PhysSuper1}
A. Blais, J. Gambetta,  \textit{et al.},
\emph{Quantum-information processing with circuit quantum electrodynamics}, 
\href{https://doi.org/10.1103/PhysRevA.75.032329}{Phys. Rev. A \textbf{75}, 032329 (2007)}.

\bibitem{PhysSuper2}
A. Wallraff, D. I. Schuster \textit{et al.},
\emph{Strong coupling of a single photon to a superconducting qubit using circuit quantum electrodynamics}, 
\href{https://doi.org/10.1038/nature02851}{Nature \textbf{431}, 162–167 (2004)}.

\bibitem{PhysSuper3}
I. Chiorescu, P. Bertet \textit{et al.},
\emph{Coherent dynamics of a flux qubit coupled to a harmonic oscillator}, 
\href{https://doi.org/10.1038/nature02831}{Nature \textbf{431}, 159–162 (2004)}.


\bibitem{PhysSolid1}
N. Samkharadze, G. Zheng \textit{et al.},
\emph{Strong spin-photon coupling in silicon}, 
\href{https://doi.org/10.1126/science.aar4054}{Science \textbf{359}, 1123-1127 (2018)}.

\bibitem{PhysSolid2}
X. Mi, M. Benito \textit{et al.},
\emph{A coherent spin–photon interface in silicon}, 
\href{https://doi.org/10.1038/nature25769}{Nature \textbf{555}, 599-603 (2018)}.

\bibitem{PhysSolid3}
A. J. Landig, J. V. Koski \textit{et al.},
\emph{Coherent spin–photon coupling using a resonant exchange qubit}, 
\href{https://www.nature.com/articles/s41586-018-0365-y}{Nature \textbf{560}, 179–184 (2018)}.



\bibitem{PhysRygberHogan}
A. A. Morgan, and S. D. Hogan,
\emph{Coupling Rydberg Atoms to Microwave Fields in a Superconducting Coplanar Waveguide Resonator}, 
\href{https://doi.org/10.1103/PhysRevLett.124.193604}{Phys. Rev. Lett. \textbf{124}, 193604 (2020)}.

\bibitem{rydberg_array}
J. D. Santis, B. Dura-Kovács \textit{et al}, \emph{Realization of a cavity-coupled Rydberg array} \href{https://arxiv.org/abs/2602.12152}{arXiv, 2602.12152 (2026)}.

\bibitem{Phys2Super1}
J. Majer, J. Chow \textit{et al.},
\emph{Coupling superconducting qubits via a cavity bus}, 
\href{https://doi.org/10.1038/nature06184}{Nature \textbf{449}, 443–447 (2007)}.

\bibitem{Pritchard2014} J. D. Pritchard, J. A. Isaacs \textit{et al.}, 
\emph{Hybrid atom-photon quantum gate in a superconducting microwave resonator}, 
\href{https://doi.org/10.1103/PhysRevA.89.010301}{Phys. Rev. A \textbf{89}, 010301 (2014)}.



\bibitem{Ruskov2003} R. Ruskov and A. N. Korotkov, 
\emph{Entanglement of solid-state qubits by measurement}, 
\href{https://doi.org/10.1103/PhysRevB.67.241305}{Phys. Rev. B \textbf{67}, 241305 (2003)}.

\bibitem{Lalumiere2010} K. Lalumière, J. M. Gambetta, and A. Blais, 
\emph{Tunable joint measurements in the dispersive regime of cavity QED}, 
\href{https://doi.org/10.1103/PhysRevA.81.040301}{Phys. Rev. A \textbf{81}, 040301 (2010)}.

\bibitem{Roch2014} N. Roch,  M. Schwartz \textit{et al.}, 
\emph{Observation of measurement-induced entanglement and quantum trajectories of remote superconducting qubits}, 
\href{https://doi.org/10.1103/PhysRevLett.112.170501}{Phys. Rev. Lett. \textbf{112}, 170501 (2014)}.

\bibitem{PhysSolidMeasureBased}
R. L. Delva, J. Mielke, G. Burkard, and J. R. Petta,
\emph{Measurement-based entanglement of semiconductor spin qubits}, 
\href{https://doi.org/10.1103/PhysRevB.110.035304}{Phys. Rev. B \textbf{110}, 035304 (2024)}.

\bibitem{delva2024} R. L. Delva, J. Mielke, G. Burkard, and J. R. Petta, 
\emph{Measurement-based entanglement of semiconductor spin qubits}, 
\href{https://doi.org/10.1103/PhysRevB.110.035304}{Phys. Rev. B \textbf{110}, 035304 (2024)}.

\bibitem{loop2} C. Budroni, G. Vitagliano, G. Colangelo, R. J. Sewell, O. Gühne, G. Tóth, and M. W. Mitchell, \emph{Quantum Nondemolition Measurement Enables Macroscopic Leggett-Garg Tests}, \href{https://doi.org/10.1103/PhysRevLett.115.200403}{Phys. Rev. Lett. \textbf{115}, 200403 (2015)}.

\bibitem{loop4} L. Rosales-Zárate, B. Opanchuk, Q. Y. He, and M. D. Reid, \emph{Leggett-Garg tests of macrorealism for bosonic systems including double-well Bose-Einstein condensates and atom interferometers}, \href{https://doi.org/10.1103/PhysRevA.97.042114}{Phys. Rev. A \textbf{97}, 042114 (2018)}.

\bibitem{loop5} G. Vitagliano, and C. Budroni, \emph{Leggett-Garg macrorealism and temporal correlations}, \href{https://doi.org/10.1103/PhysRevA.107.040101}{Phys. Rev. A \textbf{107}, 040101 (2023)}.

\bibitem{upcoming} B. Zindorf, L. Braccini, D. Das, and S. Bose, \emph{How "Quantum" is your Quantum Computer? Macrorealism-based Benchmarking via Mid-Circuit Parity Measurements}, \href{https://arxiv.org/abs/2511.15881}{arXiv:2511.15881 [quant-ph]}.





\bibitem{LGIQC1} E. Huffman, and A. Mizel, \emph{Violation of noninvasive macrorealism by a superconducting qubit: Implementation of a Leggett-Garg test that addresses the clumsiness loophole}, \href{https://doi.org/10.1103/PhysRevA.95.032131}{Phys. Rev. A \textbf{95}, 032131 (2017)}.

\bibitem{LGIQC2} H.-Y. Ku, N. Lambert, F.-J. Chan, C. Emary, Y.-N. Chen, and F. Nori,  \emph{Experimental test of non-macrorealistic cat states in the cloud}, \href{https://doi.org/10.1038/s41534-020-00321-x}{npj Quantum Inf \textbf{6}, 98 (2020)}. 

\bibitem{LGIQC3} A. Santini, and V. Vitale, \emph{Experimental violations of Leggett-Garg inequalities on a quantum computer}, \href{https://doi.org/10.1103/PhysRevA.105.032610}{Phys. Rev. A \textbf{105}, 032610 (2022)}.

\bibitem{LGIQC4} P. P. Nath, A. Sinha, and U. Sinha, \emph{Certified Random Number Generation using Quantum Computers}, \href{https://doi.org/10.48550/arXiv.2502.02973}{arXiv:2502.02973 [quant-ph]}.

\bibitem{LGIQC5} D. Melegari, M. Cardi, and P. Solinas, \emph{Quantum simulations of macrorealism violation via the QNDM protocol}, \href{https://doi.org/10.48550/arXiv.2502.17040}{arXiv:2502.17040 [quant-ph]}.


 \bibitem{JC_physics} A. D. Greentree, J. Koch, and J. Larson, \emph{Fifty years of Jaynes–Cummings physics}, \href{https://iopscience.iop.org/article/10.1088/0953-4075/46/22/220201}{J. Phys. B: At. Mol. Opt. Phys. \textbf{46} 220201}.

\bibitem{dicke_coherence_1954} R. H. Dicke,  \emph{Coherence in Spontaneous Radiation Processes}, 
  \href{https://link.aps.org/doi/10.1103/PhysRev.93.99}{Phys. Rev. \textbf{93}, 99 (1954)}.

\bibitem{schliemann_coherent_2015} J. Schliemann,  \emph{Coherent Quantum Dynamics: What Fluctuations Can Tell}, \href{https://journals.aps.org/pra/abstract/10.1103/PhysRevA.92.022108}{Phys. Rev. A \textbf{92}, 022108 (2015)}.

\bibitem{wang_giant_2022} Z. Wang, Y. Wang \textit{et al.},  \emph{Giant spin ensembles in waveguide magnonics}, \href{https://www.nature.com/articles/s41467-022-35174-9}{Nature Communications \textbf{13}, 7580 (2022)}.



\bibitem{ma_quantum_2011} J. Ma, X. Wang, C. P. Sun, and F. Nori,  \emph{Quantum spin squeezing}, \href{https://www.sciencedirect.com/science/article/pii/S0370157311002201}{Physics Reports \textbf{509}, 89-165
 (2011)}.

\bibitem{travis} M. Tavis and T. Cummings,,  \emph{Exact Solution for an $N$-Molecule---Radiation-Field Hamiltonian}, \href{https://link.aps.org/doi/10.1103/PhysRev.170.379}{Phys. Rev. \textbf{170}, 379--384
 (1968)}. 


\bibitem{ferraro_gaussian_2005} A. Ferraro,  S. Olivares, and M. G. A. Paris \emph{Gaussian states in continuous variable quantum information} \href{https://arxiv.org/abs/quant-ph/0503237}{arXiv, quant-ph/0503237, (2005)}.

    
\bibitem{schlosshauer_quantum_2007} M. Schlosshauer, \emph{Decoherence and the Quantum-To-Classical Transition}, \href{https://doi.org/10.1007/978-3-540-35775-9}{Springer International Publishing (2007)}.	

\bibitem{classicality_emergence} F. Bibak, C. Cepollaro \textit{et al.}, \emph{The classical limit of quantum mechanics through coarse-grained measurements} \href{https://arxiv.org/abs/2503.15642}{arXiv, quant-ph/2503.15642, (2025)}.



\bibitem{loop1} M. Wilde, and A. Mizel, \emph{Addressing the Clumsiness Loophole in a Leggett-Garg Test of Macrorealism}, \href{https://doi.org/10.1007/s10701-011-9598-4}{Found. Phys. {\bf 42}, 256–26 (2012)}.



\bibitem{loop3} C. Robens, W. Alt, D. Meschede, C. Emary, and A. Alberti, \emph{Ideal Negative Measurements in Quantum Walks Disprove Theories Based on Classical Trajectories}, \href{https://doi.org/10.1103/PhysRevX.5.011003}{Phys. Rev. X \textbf{5}, 011003 (2015)}.






\bibitem{SGI1} S. Bose, A. Mazumdar \textit{et al.}, \emph{Spin entanglement witness for quantum gravity}, \href{https://doi.org/10.1103/PhysRevLett.119.240401}{Phys. Rev. Lett. \textbf{119}, 240401 (2017)}.

 \bibitem{SGI2} Y. Margalit, O. Dobkowski \textit{et al.}, \emph{Realization of a complete stern-gerlach interferometer: Toward a test of quantum gravity}, \href{https://www.science.org/doi/10.1126/sciadv.abg2879}{Science Advances \textbf{7}, eabg2879 (2021)}.

 \bibitem{cat1} A. Ourjoumtsev, R. Tualle-Brouri, J. Laurat, and P. Grangier, \emph{Generating Optical Schrödinger Kittens for Quantum Information Processing.}, \href{https://www.science.org/doi/10.1126/science.1122858}{Science \textbf{312}, 83-86 (2006)}.

 \bibitem{cat2} A. Ourjoumtsev, H. Jeong,  R. Tualle-Brouri, and P. Grangier, \emph{Generation of optical ‘Schrödinger cats’ from photon number states}, \href{https://doi.org/10.1038/nature06054}{Nature \textbf{448}, 784–786 (2007)}.

\bibitem{PhysChoice}
B. Danjou, and G Burkard,
\emph{Optimal dispersive readout of a spin qubit with a microwave resonator}, 
\href{https://doi.org/10.1103/PhysRevB.100.245427}{Phys. Rev. B \textbf{100}, 245427 (2019)}.



\bibitem{Phys3Super1}
J. Kang, C. Kim, Y. Kim, and Younghun Kwon,
\emph{New design of three-qubit system with three transmons and a single fixed-frequency resonator coupler}, 
\href{https://doi.org/10.1038/s41598-025-94448-6}{Scientific Reports \textbf{15}, 12134 (2025)}.



\bibitem{transmission_lines}
R.E.Collin,
\emph{Foundation of Microwave Engineering}, 
IEEE Press Series on Electromagnetic Wave Theory.

\bibitem{CWR} M. Göppl \textit{et al.}, 
\emph{Coplanar waveguide resonators for circuit quantum electrodynamics}, 
\href{https://doi.org/10.1063/1.3010859}{J. Appl. Phys. \textbf{104}, 113904 (2008)}.





\bibitem{PhysTable4}
P. Stano, and D. Loss, 
\emph{Review of performance metrics of spin qubits in gated semiconducting nanostructures}, 
\href{https://doi.org/10.1038/s42254-022-00484-w}{Nat Rev Phys \textbf{4}, 672–688 (2022)}.




\bibitem{PhysSolid4}
X. Zhou, X. Li \textit{et al.}, 
\emph{Electron charge qubit with 0.1 millisecond coherence time}, 
\href{https://doi.org/10.1038/s41567-023-02247-5}{Nat. Phys. \textbf{20}, 116–122 (2024)}.

\bibitem{guide1} A. Megrant,  C. Neill \textit{et al.}, 
\emph{Planar superconducting resonators with internal quality factors above one million}, 
\href{https://doi.org/10.1063/1.3693409}{App. Phys. Lett. \textbf{100}, 113510 (2012)}.

\bibitem{Parity1} B. Zeng, D. L. Zhou, and L. You, \emph{Measuring the Parity of an $N$-Qubit State}, \href{https://doi.org/10.1103/PhysRevLett.95.110502}{Phys. Rev. Lett. \textbf{95}, 110502 (2005)}.

\bibitem{Parity2} A. Kolli, B. W. Lovett, S. C. Benjamin, and T. M. Stace, \emph{All-Optical Measurement-Based Quantum-Information Processing in Quantum Dots}, \href{https://doi.org/10.1103/PhysRevLett.97.250504}{Phys. Rev. Lett. \textbf{97}, 250504 (2006)}.

\bibitem{Parity3} D. Riste, M. Dukalski, C. A. Watson, G. de Lange, M. J. Tiggelman, Ya. M. Blanter, K. W. Lehnert, R. N. Schouten, L. DiCarlo, \emph{Deterministic entanglement of superconducting qubits by parity measurement and feedback}, \href{https://doi.org/10.1038/nature12513}{Nature \textbf{502}, 350-354 (2013)}.

\bibitem{Parity4} R. J. Birrittella, P. M. Alsing, and C. C. Gerry, \emph{The parity operator: Applications in quantum metrology}, \href{https://doi.org/10.1116/5.0026148}{AVS Quantum Sci. \textbf{3}, 014701 (2021)}.


\bibitem{dissipation} D. F. Walls, and G. J. Milburn,  \emph{Effect of dissipation on quantum coherence}, \href{https://link.aps.org/doi/10.1103/PhysRevA.31.2403}{Phys. Rev. A \textbf{31}, 2403, (1985)}.	







 

%












	
\end{thebibliography}
\end{document}